%% file: ms.tex
\documentclass[10pt,sort&compress,3p]{elsarticle}
\usepackage{wrapfig}

\input{header.tex}

\usepackage[english]{babel}

    \makeatletter
    \def\ps@pprintTitle{%
       \let\@oddhead\@empty
       \let\@evenhead\@empty
       \def\@oddfoot{\reset@font\hfil\thepage\hfil}
       \let\@evenfoot\@oddfoot
    }
    \makeatother

\title{Bi-Abduction for Shapes with Ordered Data
	\footnote{This is an extended version of an earlier short paper \cite{Curry:ICECCS:2019}.
				The additions over that work include: (1) greater context regarding the use of bi-abductive inference and an illustrative example showing how it would be applied; (2) a step-by-step illustration of our technique in use; (3) greater detail concerning each stage of the proof process; (4)	an extension to the technique, adding support for both binary trees and binary search trees; (5) details of a prototype implementation of the technique and (6) experimental results from applying the prototype tool to examples taken from SL-COMP.}}

\author[1]{Christopher Curry\corref{cor1}}
\author[2]{Quang Loc Le}
\address[1]{School of Computing, Engineering and Digital Technologies,
	Teesside University,
	Middlesbrough,
	Tees Valley TS1 3BX, UK}

\address[2]{University College London,
	London,
	Gower Street WC1E 6BT, UK}

\cortext[cor1]{Corresponding Author}

\begin{document}
	\renewcommand{\draft}{}
	
	\input{abstract}
	\maketitle
	
	\input{intro}
	\input{relwk}
	\input{prelim}	
	\input{illus}
	\input{norm}
	\input{matchsub}
	\input{inf}	
	\input{sound}
	\input{implement}
	\input{eval}

	\input{conc}

	\bibliographystyle{plain}
	\bibliography{bibliography.bib}
	\newpage
	\appendix
	\input{appendix_table_spec_smallfoot}

\end{document}

%% file: header.tex
\usepackage{caption}
\usepackage{subcaption}

\usepackage{graphicx}
	\graphicspath{{./images/}}
	
\usepackage{amsmath,amssymb,amsfonts}

\usepackage{textcomp}

\usepackage{xcolor}
\usepackage{color}

\usepackage{wrapfig}

\usepackage{listings}
\usepackage{bussproofs}

\usepackage{afterpage}
\usepackage{rotating}

\usepackage{multicol}
\usepackage{multirow}

\usepackage{natbib}
\newcommand{\draftNote}[1]{}
\newcommand{\draftHL}[1]{#1}

\newcommand{\locsay}[1]{}

\newcommand{\draft}{
					\renewcommand{\draftNote}[1]{\color{blue}\textit{##1}\color{black}}
					\renewcommand{\draftHL}[1]{\color{orange}##1\color{black}}
					\renewcommand{\draftCover}{\note{\begin{center}\Huge{DRAFT VERSION\\}\Huge{\today}\end{center}}}
					\renewcommand{\locsay}[1]{{\small{{\color{red} Loc: \textit{##1}}}}}
					\renewcommand{\draftBibliography}{\bibliography{../../common/bibliography.bib}}
					}
				
\newcommand{\qt}[1]{\textquotedblleft{#1}\textquotedblright}

\newcommand{\true}{\texttt{true}}
\newcommand{\false}{\texttt{false}}
\newcommand{\emp}{\texttt{emp}}
\newcommand{\nil}{\texttt{null}}

\newcommand{\conj}{\wedge}
\newcommand{\disj}{\vee}
\newcommand{\pointsto}{{\mapsto}}
\newcommand{\sepconj}{ * }

\newcommand{\mayentail}{\vdash}

\newcommand{\notentail}{\nvdash}

\newcommand{\judge}{\vartriangleright}
\renewcommand{\judge}{\mayentail}
\newcommand{\define}{{\stackrel{def}{=}}}
\newcommand{\vecVar}[1]{\overrightarrow{#1}}

\newcommand{\lt}{{<}}

\newcommand{\eq}{{=}}
\newcommand{\lteq}{{\leq}}

\newcommand{\nteq}{{\neq}}

\newcommand{\tconj}{$\wedge$~}

\newcommand{\tsepconj}{$\ast$}

\newcommand{\equ}[1]{\begin{equation}#1\end{equation}}
\newcommand{\biab}[4]{$#1 \sepconj \textcolor{blue}{[#2]} \mayentail #3 \sepconj \textcolor{red}{[#4]}$}

\def\D{\Delta}
\def\FV{{FV}}
\newcommand{\form}[1]{\ensuremath{#1}}
\newcommand{\defsym}{\ensuremath{\overset{\text{\scriptsize{def}}}{=}}
}
\newcommand{\myit}[1]{\textit{#1}}
\newtheorem{definition}{Definition}
\newtheorem{thm}{Theorem}
\newcommand{\rulename}[1]{{\bf \scriptsize [#1]}}
\newcommand{\trule}[1]{{\scriptsize #1}}
\newcommand{\sheaps}{\ensuremath{h}}
\newcommand{\sstack}{\ensuremath{s}}
\newcommand{\force}{\ensuremath{\models}}
\def\dom{\myit{dom}}

\def\iffs{\small \myit{iff~}}
\def\Store{\myit{Heaps}}
\def\Stack{\myit{Stacks}}
\def\Locations{\myit{Loc}}
\def\Var{\myit{Var}}
\def\Flds{\myit{Fields}}
\def\Val{\myit{Values}}
\newcommand{\proofsketch}[1]{{\noindent \bf
Proof Sketch\,:~}#1{\qed}}
\def\qed{\hfill\ensuremath{\square}}

%% file: abstract.tex

	\begin{abstract}
		\draftNote{Still need to rework this section. }
		Shape analysis is of great importance for the verification of the correctness and memory-safety of heap-manipulating programs, yet such analyses have been shown to be highly difficult problems.
		The integration of separation logic into shape analyses has improved the effectiveness of the techniques, but the most significant advancement in this area is bi-abductive inference.
		Enabled by separation logic, bi-abduction - a combination of abductive inference and frame inference - is the key enabler for compositional reasoning, helping to scale up verification significantly.
		Indeed, the success of bi-abduction has led to the development of Infer, the tool used daily to verify Facebook's codebase of millions of lines of code.
		However, this success currently stays largely within the shape domain. 
		To extend this impact towards the combination of shape and arithmetic domains, in this work, we present a novel one-stage bi-abductive procedure for a combination of data structures and ordering values.
		The procedure is designed in the spirit of the Unfold-and-Match paradigm where the inference is utilized to derive any mismatched portion.
		We have also implemented a prototype solver, based on the Cyclist library, and demonstrate its capabilities over a range of benchmarks from the SL-COMP competition.
		The experimental results show that our proposal shows promise for the specification inference in an automated verification of heap-manipulating programs.
	\end{abstract}
	

%% file: intro.tex

\section{Introduction}
\label{intro}
	Heap-manipulation is a powerful building block used in many real-world applications and is frequently utilised in low-level system software such as device drivers.
	While useful, heap manipulation can also be a dangerous technique, one in which relatively simple faults can cause significant numbers of issues, ranging from software crashes \cite{Yang:CAV:2008} to security vulnerabilities \cite{Szekeres:IEEE.SP:2013}.
	As a result, ensuring the correctness and safety of such programs is of great importance, yet such analyses have been shown to be highly difficult problems.
	
	In recent years, shape analysis	has emerged as a strong candidate for accomplishing such verification tasks, and with the integration of separation logic \cite{Reynolds:LICS:2002,Ishtiaq:ACM.SIGPLAN:2001} into shape analysis techniques, an increasing number of advanced and usable tools are being developed for automated reasoning over dynamically allocated heap programs.
	The success of separation logic comes from the capabilities provided by \textit{local reasoning}: a program only has an effect over the memory it accesses, it's \textit{footprint}.
	This local reasoning is enabled by two main constructions: the separating conjunction operator (\tsepconj) and the frame rule.
	While the former is used to assert the separation of parts of the heap (i.e., $F \sepconj G$ asserts that $F$ and $G$ hold in disjoint heap regions), the latter can be used to compose the analysis results of separate program parts, such as procedures.
	
	One of the most promising techniques currently utilised to automate local reasoning is \textit{bi-abductive inference} \cite{Calcagno:POPL:2009,Calcagno:NFM:2011,Le:CAV:2014}.
    Bi-abductive inference, or bi-abduction for short, is the combination of the abductive inference and frame inference techniques into a singular analysis.
	Given two formulas in separation logic, $A$ and $C$, bi-abduction aims to identify some non-trivial terms $?M$ and $?F$ such that the entailment \begin{center} $A \sepconj [?M] \mayentail C \sepconj [?F]$ \end{center} is satisfied.	
	Bi-abduction has seen a large growth in interest in recent years due to a number of useful properties of the technique.
	First, bi-abductive techniques generally have very low requirements placed upon the end-user, with many requiring only the code of the program to be analysed. 
	Second, bi-abduction is a powerful enabler of compositional analysis: an analysis in which the final output is the combination of the results of smaller analyses over components of the program, typically procedures.
	Compositional analyses provide a number of significant benefits, including high scalability, parallelisation, the ability to undertake incremental analyses and potential support for graceful failure, all highly desirable properties for an analysis technique.
	More recent work has also included advances that support the use of bi-abductive inference over near-arbitrary data structures \cite{Le:CAV:2014}, further improving the usability of the technique.
	
	However, while bi-abduction has proven to be a highly capable technique for use in program analysis, the majority of existing implementations stay within the shape domain only.
	This restriction not only limits the range of inputs that can be supported by the technique, but also carries the potential risk of imprecision when used to analyse data structures that have pure constraints as a component of their design, such as binary search trees or sorted linked lists.
	Currently, efforts are being made towards extending bi-abduction to the combined domain, allowing bi-abductive techniques to operate over general pure constraints in addition to those shape properties, with common examples of investigated pure constraints including size, ordering and content (bag) properties.
	\draftHL{As an example, the technique outlined by Trinh et al \cite{Trinh:APLAS:2013} utilises an enrichment-based approach to the identification of pure constraints, but is fully reliant upon some previous shape analysis in order to function.
	This multi-phase approach is common in the literature of combined domain verification techniques \cite{Magill:POPL:2010}, and while it has been shown to be effective, the approach is apparently not very efficient.
	Additionally, by separating the shape and pure domain, the accuracy of the final inferred specification would be heavily reliant upon the accuracy and expressiveness of the shape information determined in the initial analysis, potentially negatively affecting the quality of the overall result.}

	\draftHL{For both effectiveness and efficiency}, in this work, we present a novel one-phase bi-abduction procedure for the combination of shape and ordering properties.
    In particular, we develop a set of inference rules and a novel algorithm to search for a sequence of rule applications.
	Our search algorithm operates based on the Unfold-and-Match principle as follows: firstly, it unfolds inductive predicates, as well as subtracting heaps identified on both sides of the bi-abductive entailment.
	If this process cannot progress due to missing heap components on the left-hand side (LHS), it invokes an abductive rule to infer part of the anti-frame.
	Likewise, if the process cannot progress due to the extra heap components on the right-hand side (RHS), it invokes a procedure to infer part of the frame.
    Our system is more expressive than the bi-abduction system used in Infer \cite{Calcagno:POPL:2009} as it could deal with inductive definitions with ordering properties.
	Additionally, our system is as powerful as the combination of bi-abduction for shape domains (e.g., \cite{Calcagno:POPL:2009,Le:CAV:2014}) and pure inference for ordering properties in the arithmetic domain (e.g., \cite{Magill:POPL:2010,Trinh:APLAS:2013}).
	Moreover, the proposed proof system is proven to be sound and terminating.
	We also present a prototype implementation of the tool, developed as an instantiation of the Cyclist library \cite{Brotherston:ASPLS:2012}, and demonstrate its capabilities over a range of benchmarks from the SL-COMP competition \cite{Sighireanu:TACAS:2019}.

	\paragraph{Organization}
		The remainder of this paper is organized as follows.
		Section \ref{relwk} discusses related works.
		Section \ref{prelim} shows the syntax of separation logic formulas and formalism of the bi-abductive problem.
		Section \ref{sec.overview} describes our overall approach, supported with an illustrative example.
		Sections \ref{sec.norm}, \ref{sec.matchsub}, and \ref{sec.inf} present our main result: a bi-abduction procedure through an inference system.
		Next, Section \ref{sec.sound} discusses soundness and termination of the proposed procedure.
		Section \ref{implement} shows the implementation and the evaluation is discussed in Section \ref{sec.eval}.
		Finally, Section \ref{concl} concludes the work.

%% file: relwk.tex

\section{Related Work}
\label{relwk}
	
	There are a number of works related to our technique. 
	For classical bi-abduction, Infer \cite{Calcagno:POPL:2009} remains the most well-known and is essentially a \qt{pure} implementation of the bi-abduction technique \cite{Calcagno:POPL:2009}. Extensions have been made to the tool, but its bi-abductive capabilities remain restricted to shape properties.
	Other techniques to take advantage of bi-abductive inference include the \texttt{JaVerT} \cite{Fragoso_Santos:POPL:2019} and \texttt{Gillian} \cite{Fragoso_Santos:PREPRINT:2020} verification and testing frameworks, which utilise shape-only bi-abduction to correct identified deficiencies in specifications during a compositional analysis of real-world programs.
	In addition to these, some extended versions of the technique have also been investigated, with the second-order bi-abduction technique of Le {\em et. al.} \cite{Le:CAV:2014,Le:TACAS:2018} for general shape-only inductive predicates beyond list and binary trees. 
	
	In the area of combined domain, there is a small number of relevant works. 
	One of the earliest in this area is the work of Trinh \textit{et. al.}, described in \cite{Trinh:APLAS:2013}. 
	\draftHL{Building upon a set of shape properties obtained from some previous shape analysis, the technique extends the shape predicates with a set of pure terms representing properties such as size or order, alongside additional predicates representing relational information over the structure.
	A forwards analysis is then undertaken to generate proof obligations for the relational predicates, which are then finalised via a fixpoint analysis, producing a precondition ensuring memory safety and termination.
	One of the most developed techniques for combined domain bi-abduction, and close to our own work, is the work of Qin et al \cite{Qin:SCP:2017}.
	This technique is also based around a fixpoint analysis of the target program, abducting necessary state components in each pass, developing stronger preconditions until a fixpoint is reached.
	This stage is guided through user-defined predicates outlining the expected data structures encountered, identifying several aspects of the pure domain in these passes.
	However, a secondary bi-abductive inference is still necessary to obtain smaller components of pure information during the final abstraction phase for each iteration, and widening operations utilised to accelerate reaching the fixpoint may introduce soundness issues.
	Since it is not truly a single-stage bi-abductive method as our system is, the performance of the tool is degraded by the two-phase analysis.}
	
	A range of other combined domain analysis techniques exist in the literature: \texttt{Thor} \cite{Magill:POPL:2010} is capable of handling pure properties alongside memory safety, able to operate over ordering, shape, size or depth properties via an additional analysis over a "proof program".
	Chang et al \cite{Chang:POPL:2008} describe a generalised framework for shape analysis based around forward abstract interpretation with additional support for relations between data values.
	The venerable TVLA system \cite{Sagiv:TOPLAS:2002} is capable of supporting a wide range of structures and properties, including complex pure properties such as ordering, though is limited by the lack of support for compositional reasoning.
	Finally, techniques based on Forest Automata are also known \cite{Holik:TACAS:2015}, though few can operate in the combined domain \cite{Abdulla:ATVA:2013}.
	

%% file: prelim.tex


\section{Preliminaries}
	\label{prelim}
	\input{prelim_fig_lang.tex}

	In this section, we present the syntax of separation logic formulas (Subsection \ref{sec.prelim.syntax}), their semantics (Subsection \ref{semantics}), and the problem targeted by our technique (Subsection \ref{sec.prelim.prob}).
	
	\subsection{Syntax}\label{sec.prelim.syntax}
	The language used in our work is shown in Figure \ref{grammar}.
	Program variables are defined as $italic$ characters, and refer to variables originating from within the program itself.
	Logical variables are indicated with upper-case letters, such as $X$, and refer to variables that appear in the analysis only.
	For readability, we often omit record fields from the notation where unambiguous: $x \pointsto [n{:}y,v{:}z]$ may be shortened to $x \pointsto[y,z]$, as an example.
	We may additionally omit the square brackets around single-field records, as in $x \pointsto y$.
	
	Our language includes inductive definitions describing fundamental list structures: $ls$, representing a simple singly-linked list, and $sls$, representing a \textit{sorted} singly-linked list.
	We have also included definitions for a binary tree $tree$ and a binary search tree $stree$, an addition to our initial work \cite{Curry:ICECCS:2019}.
	The full inductive definitions of these shape predicates may be found in Figure \ref{fig.lang.defs}.
	
	A singly-linked list $ls(E_1,E_2)$ 
        consists of a series of nodes linked via a pointer, starting with $E_1$ and ending at $E_2$ (non-inclusive).
	Each of the list predicates included in our system can be used to define a list segment or a full null-terminated list, dependant only on the value of $E_2$.
	A \textit{sorted} singly-linked list $sls(E_1,V_1,V_2,E_2)$ is a sequence of singly-linked nodes, beginning with node $E_1$ and ending at some node $E_2$ (non-inclusive), with all values stored in those nodes obeying an ascending order.
	In order to simplify the checking of these structures, the minimum and maximum values of the list are also tracked inside the predicate, with the minimum value being represented as $V_1$ and the maximum $V_2$.
	Note that $V_2$ \textit{is not} the value of node $E_2$; rather, $V_2$ would refer to the value of the final node in the list, which points to $E_2$.
	Both $tree(E)$ and $stree(E,V_1,V_2)$ describe a binary tree: a single node pointing to two known subtrees, or a \texttt{null}  root describing an empty tree.
	As with $ls$ and $sls$, the primary difference between $tree$ and $stree$ is the consistent ordering constraints present in $stree$, which ensures that all values in the left subtree of an $stree$ are lower than the value of the root node, and all values in the right subtree are greater than the value of the root.
	Minimum and maximum values are recorded in $stree$ predicates in a similar manner to $sls$ predicates.
	
	\subsection{Semantics}
	\label{semantics}
	The semantics of this fragment are quite standard, following from the semantics of separation logic with general inductive definitions and arithmetic presented in \cite{Le:CAV:2017}.
	These semantics are given by a relation \form{\sstack{,}\sheaps ~{\force}~ H} that forces the stack \form{\sstack} and heap \form{\sheaps} to satisfy the constraint \form{H}, where \form{\sheaps \in {\Store} }, \form{\sstack \in {\Stack}}, and \form{H} is a formula. Stack and heap abstractions are defined as:
	\[
	\begin{array}{lcl}
	{\Store}  & {\defsym} &  {\Val} {\rightharpoonup_{fin}} (\Flds \times \Val)^N  \\
	{\Stack} & {\defsym} &  {\Var} ~{\rightarrow}~ \Val
	\end{array}
	\]
	where \form{N} is the maximum number of fields.

	The semantics of our fragment are as follows:
	\[
	\begin{array}{lcl}
	\form{\sstack},\form{\sheaps} \force \emp &
	\iffs & \dom({\sheaps}) {=} \{\}\\
	
	\form{\sstack},\form{\sheaps} \force E{\pointsto}[f_i{:}v_i] &
	\iffs &  {\dom}(\sheaps) {=}\{\sstack(E)\} \sheaps(\sstack(E)){=} (~(f_1,\sstack({E_1})), ..., (f_N,\sstack({E_N}))~)\\
		
	\form{\sstack},\form{\sheaps} \force  \Sigma_1 \sepconj \Sigma_2 &
	\iffs & \exists \sheaps_1,\sheaps_2 {\cdot}~ \sheaps_1 {\#} \sheaps_2 \mbox{ and }
	\sheaps{=}\sheaps_1 {\cdot} \sheaps_2, 
		 \sstack,\sheaps_1 \force \Sigma_1 \mbox{ and } \sstack,\sheaps_2 \force \Sigma_2\\
		
	\form{\sstack},\form{\sheaps} \force \true &
	\iffs & \mbox{\textit{always}} \\
	
	\form{\sstack},\form{\sheaps} \force\Pi \conj \Sigma &
	\iffs &  {\sstack},\form{\sheaps} \force {\Sigma}  \text{ and }	{\sstack} \force {\Pi} \\

        \form{\sstack},\form{\sheaps} \force \exists {v} {.}\Delta &
	\iffs & \exists {\alpha} \cdot {\sstack}[v \pointsto {\alpha}],\form{\sheaps} \force {\Delta} \\

	\form{\sstack},\form{\sheaps} \force  H_1 \vee H_2 &
	\iffs & \sstack,\sheaps \force H_1 \mbox{ or } \sstack,\sheaps \force H_2\\
	\end{array}
	\]
	where \form{dom(f)} is the domain of function \form{f}, \form{\sheaps_1 {\#} \sheaps_2} denotes disjoint heaps $h_1$ and $h_2$ i.e., \form{{\dom}(\sheaps_1) {\cap} {\dom}(\sheaps_2) {=} \emptyset}, and \form{\sheaps_1 {\cdot} \sheaps_2} denotes the union of two disjoint heaps.
	If \form{\sstack} is a stack, \form{v {\in} \Var}, \form{v {\not\in} \dom(\sstack)} and \form{\alpha {\in} \Val {\cup} \Locations}, we write \form{ {\sstack}[v {\pointsto} {\alpha}] \equiv {\sstack} {\cup} \{(v, \alpha)\}}.
	Note that in a concrete memory model such as the RAM model, the field names of points-to predicates are transformed into pointer offsets.
	The pair \form{(f_i,\sstack({E_i})} (for all \form{i \in \{1...N\}}) would then be interpreted as \form{\sstack(E)} {+} \textit{off}\form{_{f_i} = \sstack({E_i})}, where \textit{off}\form{_{f_i}} is the corresponding offset of field \form{f_i}.
	The entailment \form{H \models H'} holds iff for all \form{\sstack} and \form{\sheaps}, we have if \form{\sstack,\sheaps \force H} then \form{\sstack, \sheaps \force H'}.
	Note that we also preserve \form{\sstack({\nil})} as a special value such that it is not in any domain of heaps.

	\input{prelim-biad}


%% file: prelim_fig_lang.tex
\begin{figure}[t]
	\centering
	\begin{subfigure}[b]{\textwidth}
		\centering
		\begin{tabular} {l c}
			$x,y,... \in Var$ & variables \\
            $f, f_i, ... \in Fields$ & fields \\
			$E::=\ \nil\ \vert\ x$ & expressions \\
			$V::=$ i, j, ... $\in Values$ & values \\ 
			$P::= E \eq E\  \vert\  E \lt E\ $ & simple pure formulae\\
			$\Pi ::= \true\ \vert\ P\ \vert\ \neg \Pi\ \vert\ $ $\exists v. \Pi\  \vert\ \Pi \conj \Pi$ & pure formulae \\
		    $\rho::= f_1:E_1, ..., f_k:E_k$ & record expressions \\
			$\Sigma::= \emp\ \vert\ E\pointsto[\rho]  \vert\ \Sigma * \Sigma\ \vert\ $  $ls(E, E)\ \vert\ sls(E,x,y,E)$\\
			$\Delta::= \Pi \conj \Sigma$ & qf symbolic heaps \\ 
			$H::= \exists\vecVar{X}.\Delta$ & symbolic heaps \\
		\end{tabular}
		\caption{Grammar Definitions}
		\vspace{2ex}
		\label{grammar}
		\label{fig.lang.grammar}
	\end{subfigure}
	\hfill
	\vspace{5pt}
	\begin{subfigure}[b]{\textwidth}
		\centering \small
		\begin{tabular} {lcl}
			$ls(E_1,E_2)$ & $\define$ & $E_1 = E_2 \conj \emp $  $ \disj~ \exists E'.E_1 \neq E_2 \conj E_1\pointsto[E'] \sepconj ls(E',E_2)$ \\
			$sls(E_1,V_1,V_2,E_2)$ & $\define$ & $E_1 {=} E_2 \conj V_1 {=} V_2 \conj \emp\ $ $\disj~\exists E',V'.V_1{\leq} V' \conj E_1 {\neq} E_2 \conj E_1\pointsto[E',V_1] \sepconj  sls(E',V',V_2,E_2)$ \\
			$tree(E)$ & $\define$ & $ E=null \conj \emp \disj E\pointsto[l,r] \sepconj tree(l) \sepconj tree(r)$\\
			$stree(E,V_1,V_2)$ & $\define$ & $ E=null \conj V_1 = V_2 \conj \emp$
                         \\ && \quad $\disj~ \exists V_3,V_4. V_3\lteq V' \conj V' \lteq V_4 \conj E\pointsto[l,r,V'] \sepconj stree(l,V_1,V_3) \sepconj stree(r, V_4, V_2)$\\
		\end{tabular}
		\caption{Predicate Definitions}
		\label{fig.lang.defs}
	\end{subfigure}
	\caption{Language Fragment}
	\label{fig.lang}
\end{figure}

%% file: prelim-biad.tex
	\subsection{Bi-Abduction}\label{sec.prelim.prob}
		
		In this subsection, we first discuss two sub-problems: abductive inference and frame inference. Following this, we formally define the bi-abductive problem.

		\paragraph{Abductive Inference}	
			Abductive inference is an analysis technique that aims to identify the \qt{missing} portions of a program state required to satisfy some entailment. Initially inspired by the work of Peirce \cite{Peirce:1965}, abductive inference has seen a surge in interest over recent years, with its applications in the field of shape analysis being the focus of this work. In general, abductive inference aims to find some \textit{anti-frame} $?M$ such that the entailment: \form{A \sepconj [?M] \mayentail C} is satisfied\footnote{A variation of abductive inference for classical logic is also described in the literature; it is mechanically identical, using classical logic's \tconj in place of separation logic's \tsepconj.}.
				
			For a given entailment, there may be many possible solutions for the anti-frame, often varying in the level of detail outlined by a particular formula.
			As a result of this, abductive inference techniques often make use of a ranking function in order to identify the best solution for a given abduction.
			These ranking functions vary between implementations, but typically favour the solution with the smallest spatial footprint in order to generate higher-quality specifications.
			Additional constraints are also enforced over the possible solutions: trivial solutions such as $?M = \false$ are prohibited in order to ensure a useful solution is identified.

		\paragraph{Frame Inference}
			Frame inference can be seen as the counterpart to abductive inference. Given an entailment $A \mayentail C$, frame inference aims to identify some $?F$ such that $A \mayentail C\ \sepconj\ [?F]$ holds. This $?F$ is the \textit{frame} of the entailment, some additional fragment of the program state that is not necessary to prove the entailment. For example, consider the entailment:
				\form{x\neq y \wedge  x\pointsto3 \sepconj y \pointsto4 \mayentail x\pointsto3 \sepconj [?F]}.
			It can be readily seen that this entailment can be satisfied, as $x\pointsto3$ is present in both the antecedent and consequent. However, the fragment $y \pointsto4$ is not necessary to show the satisfiability of the entailment and could be removed without a negative effect; $y \pointsto4$ is the frame of this example. 
				
			The identification of the frame of an entailment in an analysis is crucial to supporting both the use of procedure specifications and compositional analyses. By identifying the portion of the state that will not be modified by a given procedure, the postcondition of the invoked procedure, when extended with the frame, will be an accurate representation of the program state upon return. This extended postcondition can be directly adopted as the state, allowing for a simple continuation of the analysis without the need to analyse the procedure itself, assuming the specification is known.

		\paragraph{Bi-Abduction}
			We use \form{\FV(H)} to denote all free variables in formula \form{H}.
			Throughout this work, we consider the bi-abductive problem which is defined as:
			\[
			\begin{array}{|ll|}
			\hline
			 \quad \text{PROBLEM:} & {{\tt QF{\_}BIABD}}.  \\
			 \quad \text{INPUT:} & \form{\D} \text{ and } \form{\D'}\text{ where }  \FV(\D') \subseteq \FV(\D)\cup\{\nil\}. \\
			 \quad \text{QUESTION:} & \text{Does there exist } \form{?M}, \form{?F} ~such~ that~ \form{\D \sepconj {?M} ~\models~ \D' \sepconj {?F} } \text{ holds? }  \\
			\hline
			\end{array}
			\]

%% file: illus.tex
\newcommand{\cequ}[1]{\begin{equation}#1\end{equation}}
\section{Overview and Illustration}

\label{sec.overview}
\input{mov}

\subsection{A Bi-Abduction Example}
\label{illustration}
	We will now demonstrate our technique over an example entailment.
	This example, shown below, features \textit{sorted} singly linked lists, with a known root and (non-inclusive) end-point, as well as a known minimum and maximum value.
	This predicate, discussed in detail in Section \ref{prelim}, can be seen in full in Figure \ref{fig.lang.defs}.
	\equ{w\pointsto[x,i] \sepconj x\pointsto[y,j] \sepconj sls(y,k,l,z) \sepconj [?M] \mayentail sls(x,j,l,z) \sepconj z\pointsto\nil \sepconj [?F]}
	
	Our technique aims to identify whether a given entailment holds, and if not, identify some values for the symbolic heap fragments $?M$ and $?F$.
	These fragments, named the \textit{anti-frame} and \textit{frame} respectively, when introduced into the initial invalid entailment, will ensure that the now corrected entailment will hold (note that we will omit these placeholders for spacing reasons in this example).
	In order to achieve this, our technique first aims to convert the entailment into an appropriate normalized form.
	In this form, predicates in the antecedent will have a singular configuration, ensuring there will be very little ambiguity as to the contents of the program memory.
	The initial stage of this process is to identify implicit constraints and introduce pure terms in order to ensure that those constraints are explicit in the entailment. 
	Initially, this will produce the entailment:	
	\cequ{w\nteq\nil \conj x\nteq\nil \conj w\nteq x \conj w\pointsto[x,i] \sepconj x\pointsto[y,j] \sepconj sls(y,k,l,z) \mayentail sls(x,j,l,z) \sepconj z\pointsto\nil}
	
	In this entailment, the two assigned nodes, $w$ and $x$ now have explicit non-null constraints, as any assigned node cannot by definition be null.
	Similarly, as there are now two separate, non-null, non-empty spatial terms, these two variables cannot be aliases, and so an inequality between the two is also introduced.
	These terms ensure that certain contradictory inferences are detected, or in some cases prevented, and ensuring that invalid branches do not progress.
	
	Following this, the next key part of this transformation is to force any shape predicates (in this case, the \textit{sls} predicate) into either the base case, indicating that the predicate is equivalent to an empty heap, or the recursive case, ensuring that the predicate refers to at least a single assigned node in memory.
	In order to achieve this, we apply a form of the excluded middle rule, identifying equalities and disequalities such that the entailment will only permit one of the two cases of the shape predicate.
	This process will also create branches in the entailment, one for each configuration of the shape predicate, allowing for the exploration of all possible states of the data structure.
	For this example, the two produced branches are as follows:
	\cequ{w\nteq\nil \conj x\nteq\nil \conj x\nteq w \conj y\nteq z \conj w\pointsto[x,i] \sepconj x\pointsto[y,j] \sepconj sls(y,k,l,z) \mayentail sls(x,j,l,z) \sepconj z\pointsto\nil}
	\cequ{w\nteq\nil \conj x\nteq\nil \conj x\nteq w \conj y{=}z \conj w\pointsto[x,i] \sepconj x\pointsto[y,j] \sepconj sls(y,k,l,z) \mayentail sls(x,j,l,z) \sepconj z\pointsto\nil}

	Entailment 3 represents the case in which the sorted list is non-empty, and Entailment 4 represents the case in which the predicate is empty.
	The inclusion of the equality (or inequality) between $y$ and $z$ ensures that, when expanded, the predicate definition can only permit one of the cases to hold, thus eliminating the other from that branch of the entailment.
	From this point, the illustration focuses on Entailment 3, the recursive case.
	The technique proceeds onwards, identifying a small number of further constraints:
	first, as the $sls$ predicate is non-empty, $y$, the head of that list must be non-null.
	Secondly, as in the earlier stages, constraints enforcing the non-aliasing of $y$ with the other non-null variables are identified, and finally, in order to ease further inference, the ordering constraint between the minimum and maximum values in the predicate is also expressed as a separate constraint.
	Introducing these constraints into the entailment concludes the normalisation phase, as the resulting antecedent can now only refer to a single heap configuration.
	The fully normalised entailment is as follows:
	\begin{align}
	  &w\nteq\nil \conj x\nteq\nil \conj y\nteq\nil\conj x\nteq w \conj y\nteq w \conj y\nteq x \conj y\nteq z \conj k\lteq l \conj  w\pointsto[x,i] \sepconj x\pointsto[y,j] \sepconj sls(y,k,l,z) \nonumber\\
	 &\qquad \mayentail sls(x,j,l,z) \sepconj z\pointsto\nil
	\end{align}
	
	Once no further normalisation can take place, the system will begin to match and subtract spatial components of the entailment, aiming to reduce the entailment to an empty heap.
	Initially, the procedure will match the node $x\pointsto[y,j]$ against the $sls$ predicate in the consequent, recognising that they share a head variable.
	The procedure will then implicitly unfold the predicate, match the concrete node with the freshly concretized node in the consequent and subtract both from the entailment.
	\begin{align}
	 &w\nteq\nil \conj x\nteq\nil \conj y\nteq\nil\conj x\nteq w \conj y\nteq w \conj y\nteq x \conj y\nteq z \conj k\lteq l \conj  w\pointsto[x,i] \sepconj sls(y,k,l,z) \nonumber\\
	&\qquad \mayentail j\lteq k \conj sls(y,k,l,z) \sepconj z\pointsto\nil
	\end{align}
	
	The procedure would also repeat any normalisation steps at this stage, but no such steps are needed in this example.
	Following this, another match is identified, recognising that there are identical spatial terms on each side of the entailment: $sls(y,k,l,z)$.
	As the predicates are identical, there is no need to unfold either of the predicates, and so both terms are simply removed, leaving the entailment	
	\equ{w\nteq\nil \conj x\nteq\nil \conj y\nteq\nil\conj x\nteq w \conj y\nteq w \conj y\nteq x \conj y\nteq z \conj k\lteq l \conj  w\pointsto[x,i] \mayentail j\lteq k \conj z\pointsto\nil}
	
	At this stage, there are no further matches that can be made with the remaining terms, both spatial and pure.
	As a result, the procedure continues on to attempting to infer some aspect of the antecedent or consequent that would allow for the proof to continue.
	In this case, the only solution is to recognise that the remaining spatial nodes are part of the frame and anti-frame.
	To handle this, the procedure applies inference rules that record the relevant terms as part of either the frame or anti-frame, and removes them from the entailment.
	In this manner, the remaining nodes are subtracted from the entailment, and are tracked separately as part of either $?M$ or $?F$.
	After applying these rules, only pure terms remain:
	\equ{w\nteq\nil \conj x\nteq\nil \conj y\nteq\nil\conj x\nteq w \conj y\nteq w \conj y\nteq x \conj y\nteq z \conj k\lteq l \mayentail j\lteq k}
	
	As there is no overlap between these pure terms, and no further refinement or subtractions can be made, the procedure concludes that these pure terms are part of the frame and anti-frame.
	Applying a final inference rule, all remaining pure terms are recoreded as part of the relevant inferred fragment and subtracted from the entailment, reducing it to $\true \conj \emp \mayentail \true \conj \emp$ and ending the proof of this branch.
	
	From here, the procedure backtracks to the second branch produced by the case-split over the predicate and repeats this process.
	The proof proceeds in a highly similar manner, with one significant exception: immediately following the branching caused by splitting the entailment, the produced equality is applied to the whole entailment.
	The result:
	\equ{w\nteq\nil \conj x\nteq\nil \conj x\nteq w \conj w\pointsto[x,i] \sepconj x\pointsto[y,j] \sepconj sls(y,k,l,y) \mayentail sls(x,j,l,y) \sepconj z\pointsto\nil}
	includes an $sls$ predicate where both the start and end point are identical.
	In this case, the predicate equates to \emp, causing the procedure will subtract the predicate and introduce an equality over the values in order to preserve that information.
	This eliminates both list fragments much earlier in the proof, accelerating the eventual conclusion of the branch.
	
	Once all branches in the proof have been completed, the procedure collects all fragments of the anti-frame and propagates them back to the initial entailment, and  gathers the final resulting frame from each branch in a similar way.
	Once substituted back into the initial entailment, we obtain the final outputs:
	\[
		\begin{array}{l}
			z\nteq\nil\conj j\lteq k \conj w\pointsto[x,i]\sepconj x\pointsto[y,j] \pointsto sls(y,k,l,z) \sepconj z\pointsto\nil\\
			\qquad  \mayentail w\nteq\nil \conj x\nteq\nil \conj y\nteq\nil\conj x\nteq w \conj y\nteq w \conj y\nteq x \conj y\nteq z \conj k\lteq l \conj w\pointsto[x,i] \sepconj sls(x,j,l,z) \sepconj z\pointsto\nil \\
			\\
         	l{=}k \conj y{=}z \conj y\nteq\nil \conj j\lteq l \sepconj w\pointsto[x,i] \sepconj x\pointsto[y,j] \sepconj z\pointsto\nil \sepconj sls(y,k,l,z) \\
         	\qquad \mayentail l{=}k \conj y{=}z \conj x\nteq\nil \conj w\nteq\nil \conj x\nteq w \conj w\pointsto[x,i] \sepconj sls(x,j,l,z) \sepconj z\pointsto\nil
		\end{array} 
	\]

	In the following sections, we will go into further details as to the purpose of each stage of our procedure, as well as the the design and utilisation of the inductive rules of those stages.
	For the normalisation stage, Section \ref{sec.norm} describes its purpose in greater detail, as well as presenting the rules employed to transform entailments into that normalised form.
	For the core matching and subtraction stage, Section \ref{sec.matchsub} outlines the rules, aims and processes of that step of our procedure.
	Finally, Section \ref{sec.inf} presents our inference rules, and discusses both their use and design intentions.
		

%% file: mov.tex
\draftNote{Some refinements can be made here. Need to rework the program-based example and discussion to demonstrate that this is a more general approach that can benefit from biabduction, rather than something we do ourselves. Some slight reworks and refinements to the rest of the content are possible, but as this is taken from previous paper, not sure there is much to do there.}

\begin{figure}[tb]
\begin{lstlisting}
Node insertSorted(Node l, Node y){
//Inf. PRE: sls(l,I,J,null) $\sepconj$ y$\pointsto$[N,K]
    Node p = findPos(l, y.val);
    if (p == null) {
      y.next = l;
      return y;
    }
    else {
      y.next = p.next;
      p.next = y;
      return l;
    }
}//Inf. POST: 
//ret=y $\conj$ K<I $\conj$ y$\pointsto$[l,K] $\sepconj$ sls(l,I,J,null) $\disj$
//ret=l $\conj$ p$\nteq$null $\conj$ A$\lteq$B $\conj$ B<K $\conj$ B$\lteq$C $\conj$ K$\lteq$C $\conj$
// sls(l,I,A,p) $\sepconj$ p$\pointsto$[y,B] $\sepconj$ y$\pointsto$[M,K] $\sepconj$ sls(M,C,D,null)

Node findPos(Node l, int x){   
//SUMMARY ONLY
//Given PRE: 
//sls(l,i,j,null);
//Given POST:
//ret=null$\conj$x<i$\conj$sls(l,i,j,null) $\disj$ 
// A$\lteq$B $\conj$ B<x $\conj$ B$\lteq$C $\conj$ x$\lteq$C $\conj$
// sls(l,i,A,ret) $\sepconj$ ret$\pointsto$[M,B] $\sepconj$ sls(M,C,D,null)
}
\end{lstlisting}
	\caption{Sorted List Insertion}
	\label{mov:fig:sls_ins}
\end{figure}

To illustrate how bi-abduction is applied for specification inference, consider the program detailed in Figure \ref{mov:fig:sls_ins}.
\texttt{insertSorted} outlines a basic insertion procedure for sorted lists. It is designed to insert the node \texttt{y} into the sorted list rooted at \texttt{l}, preserving the ordering properties of the data structure.
Assuming we are given a procedure summary (a specification using Hoare triples) of the procedure \texttt{findPos}, we show how to infer the pre and post of \texttt{insertSorted} at lines 2 and 13.

The inference begins with the precondition \form{\emp} and proceeds to execute the code at line 3.
In order to show that the procedure does not contain a null-dereference error over pointer \form{y}, we call the bi-abductive procedure with the following query:
\[
   \emp \sepconj [?M] \judge \exists N,K.~ y\pointsto[N,K] \sepconj [?F]
\]
The procedure identifies an anti-frame \form{?M\equiv y\pointsto[N,K]} and the corresponding frame \form{?F\equiv \emp}, leading us to infer that we would execute the procedure with precondition $y\pointsto[N,K]$ rather than $\emp$.

We then proceed to the procedure call for \texttt{findPos}, using its summary.
By means of compositional analysis, we send the following query to the bi-abductive procedure:
\[
    y\pointsto[N,K] \sepconj [?M] \judge \exists I,J.~ sls(l,I,J,null) \sepconj [?F]
\]
which returns \form{?M\equiv \exists I,J.~ sls(l,I,J,null)} and \form{?F\equiv y\pointsto[N,K]}.
This inference identifies the fact that in order to satisfy the precondition of \texttt{findPos}, \texttt{l} must be the head of a sorted list and \texttt{y} must be an assigned node with a value field, a state that can intuitively be seen as reasonable.
While the anti-frame is spatially conjoined into the \texttt{insertSorted} precondition, the frame is composed with the post of \texttt{findPos} to constitute the post-state of the call.
We then continue the execution forward until we obtain the postconditions of the \texttt{return} statements, and so the procedure.

\subsection{Bi-Abductive Procedure}
The core of the proposed bi-abductive procedure is a search algorithm that, given some starting entailment, scans through the set of proof rules for a rule that can be applied on the entailment.
Once such a rule is identified, it is applied to the entailment to obtain some new sub-goals (and possibly fragments of the frame or anti-frame) until it can make a decision as to whether the entailment holds. 
All proof rules in our system are checked and applied in a specific order, aiming to ensure the process identifies the weakest possible preconditions and the strongest possible postconditions, while maintaining the validity of the solutions and forbidding trivial or \false\ preconditions. The rules are searched in the following order:

\begin{enumerate}
	\item Firstly, rules in Sect. \ref{sec.norm} are applied exhaustively to unfold all possible scenarios and normalize
	the antecedent of the input entailment.
	\item Secondly, rules in Sect. \ref{sec.matchsub} are applied to match  terms across the antecedent and consequent, removing them where possible.
	If heaps in the two sides are empty, the algorithm returns successfully.
	\item Lastly, it checks whether or not any rules in Sect. \ref{sec.inf} could be applied to infer an anti-frame and frame.
	That is to identify portions of heap and pure information missing/immutable from the entailment, moving terms missing from the antecedent to the precondition and moving immutable terms from the antecedent to the postconditions.
	After that, it repeats the first phase.
\end{enumerate}

The normalization in the first phase is essential for the subsequent two phases.
It ensures that every spatial predicate in a normalized formula is precise; that is, given that \form{\sstack, \sheaps \force H} and H is in normal form, then for any spatial predicate \form{p \in H}, there exists {\em one and only one} sub-heap \form{\sheaps' \subseteq \sheaps} such that \form{\sstack, \sheaps' \force p}. 
In turn, this precision helps to guide the matching and inference rules, as given a predicate \form{p} on one side of a normalized entailment, there exists one and only one predicate \form{p'} on the other side such that \form{p} and \form{p'} are matched using the rules in the second phase.
Furthermore, if we could not find such a predicate \form{p'} in the existing entailment, \form{p'} is eventually missing.
The inference rules in the third phase reveal the missing predicates and then add them back to the entailment.
This forms the foundation for the inference mechanism in our work.

%% file: norm.tex
\section{Unfolding and Normalization} \label{sec.norm}
\begin{figure*}[tb]
\begin{center}
\[
	\begin{array}{c}
		\rulename{SUBST}\\
		\AxiomC{$(\Pi \conj \Sigma)[E/x] \sepconj [M] \judge \D[E/x]  \sepconj [F]$}
		\UnaryInfC{$(\Pi \conj x{=}E \conj \Sigma) \sepconj [M\conj x{=}E] \judge \D \sepconj [F\conj x{=}E]$}
		\DisplayProof
	\end{array}
	\quad
	\begin{array}{c}
		\rulename{LIDENT=}\\
		\AxiomC{$(\D) \sepconj [M] \judge \D'\sepconj [F]$}
        \UnaryInfC{$(\D \conj E{=}E) \sepconj [M] \judge \D'\sepconj [F]$}
		\DisplayProof
	\end{array}
\]
\[
	\begin{array}{c}
		\rulename{EXCLUDE-MIDDLE}\\
		\AxiomC{$(E_1{=}E_2 \conj \D_1) \sepconj [M_1] \judge \D'\sepconj [F_1]$}
	    \AxiomC{$(E_1{\neq}E_2 \conj \D_2) \sepconj [M_2] \judge \D'\sepconj [F_2]$}
		\BinaryInfC{$\D \sepconj [M_1\conj M_2] \judge \D'\sepconj [F_1 \vee F_2]$}
		\DisplayProof\\
		(\text{where } \form{\FV({E_1,E_2}) \subseteq (\FV(\D_1) {\cup} \FV(\D_2))})
	\end{array}
\]
\[
	\begin{array}{c}
		\rulename{LS{-}LBASE}\\
		\AxiomC{$(\D) \sepconj [M] \judge \D'\sepconj [F]$}
		\UnaryInfC{$(\D \sepconj ls(E,E)) \sepconj [M] \judge \D'\sepconj [F]$}
		\DisplayProof
	\end{array}
	\quad
	\begin{array}{c}
		\rulename{SLS{-}LBASE}\\
		\AxiomC{$(\D[V'/V]) \sepconj [M] \judge \D'[V'/V]\sepconj [F]$}
		\UnaryInfC{$(\D \sepconj sls(E,V,V',E)) \sepconj [M \conj V=V'] \judge \D'\sepconj [F\conj V=V']$}
		\DisplayProof
	\end{array}
\]
\[
	\begin{array}{c}
		\rulename{TREE-LBASE}\\
		\AxiomC{$(\D) \sepconj [M] \judge \D'\sepconj [F]$}
		\UnaryInfC{$(\D \sepconj tree(\nil)) \sepconj [M] \judge \D'\sepconj [F]$}
		\DisplayProof
	\end{array}
	\quad
	\begin{array}{c}
		\rulename{STREE-LBASE}\\
		\AxiomC{$(\D[V'/V]) \sepconj [M] \judge \D'[V'/V]\sepconj [F]$}
		\UnaryInfC{$(\D \sepconj stree(\nil,V,V')) \sepconj [M \conj V{=}V'] \judge \D'\sepconj [F \conj V{=}V']$}
	\DisplayProof
	\end{array}
\]
\[
	\begin{array}{c}
		\rulename{NODE{-}EX}\\
		\AxiomC{$(G(op(E)) \conj E{\neq}\nil \conj \D \sepconj op(E)) \sepconj [M] \judge \D'\sepconj [F]$}
		\UnaryInfC{$(G(op(E)) \conj \D \sepconj op(E)) \sepconj [M] \judge \D'\sepconj [F]$}
		\DisplayProof
		\\(\text{where } \form{E{\neq}\nil {\notin} \D})
	\end{array}
	\begin{array}{c}
		\rulename{NODES{-}EX}\\
		\AxiomC{$ E_1{\neq}E_2 \conj \D \sepconj op_1(E_1){\sepconj}op_2(E_2)) \sepconj [M] \judge \D'\sepconj [F]$}
		\UnaryInfC{$(\D \sepconj op_1(E_1){\sepconj}op_2(E_2)) \sepconj [M] \judge \D'\sepconj [F]$}
		\DisplayProof \\
		(\text{where } \form{E_1{\neq}E_2 {\not\in}\D
		 \text{ and } {G(op_1(E_1))}, {G(op_2(E_2)) \in \D}})
	\end{array}
\]
\caption{Normalization Rules for Bi-Abduction}
\label{fig.ent.norm}
\end{center}
\end{figure*}

In this section, we will discuss the design of our normal form, as well as the design and usage of the corresponding normalisation rules, which are detailed in Figure \ref{fig.ent.norm}.
In the definition of our normal form, we utilise a small number of shorthand symbols to represent a range of values:
the symbol $op(E)$ denotes one of the separation logic expressions supported by our system, namely $E\pointsto[\rho]$, $ls(E,F)$, $sls(E,V,V',F)$, $tree(E)$ and $stree(E,V,V')$.
Additionally, the guard $G(op(E))$ is defined as $G(E\pointsto[\rho]) \defsym \true $, $G(ls(E,F)) \defsym  E{\neq}F $, $G(sls(E,V,V',F)) \defsym  E{\neq}F $, $G(tree(E)) \defsym E \neq \nil $, $G(stree(E,V,V')) \defsym E \neq \nil $, and describes the condition that will enforce non-empty properties over the corresponding $op$.
The conditions of the normal form itself are defined as follows:
\begin{definition}[Normal Form]\label{defn.nf}
 A formula $\Pi{\wedge}\Sigma$ is in normal form (NF for short) if:
	\begin{enumerate}
	\item $\form{op(E) \in \Sigma}$ implies $\form{G(op(E)) \in \Pi}$.
	\item $\form{op(E) \in \Sigma}$ and  $\form{G(op(E)) \in \Pi}$ imply $\form{E{\neq}\nil \in \Pi}$.
	\item $\form{op_1(E_1)\sepconj op_2(E_2) \in \Sigma}$,  $\form{G(op_1(E_1)) \in \Pi}$, and  $\form{G(op_2(E_2)) \in \Pi}$ imply
	 $\form{E_1{\neq}E_2\in \Pi}$.
	\item $\form{E_1{=}E_2 \not\in \Pi}$.
	\item $\form{E{\neq}E  \not\in \Pi}$.
	\item $\form{\Pi}$ is satisfiable.
	\end{enumerate}
\end{definition}

If \form{\D} is in NF and for any \form{\sstack} and \form{\sheaps} such that \form{\sstack, \sheaps \models \D}, \form{\dom(\sheaps)} is uniquely defined by \form{\sstack}.
A bi-abductive entailment is in NF if its antecedent is in NF.
\draftNote{Swap/move the normal form discussion/description from preliminaries to here?}

Figure \ref{fig.ent.norm} shows the inference rules used by our system to transform a bi-abductive entailment into it's NF.
For the first condition of NF, we apply \trule{EXCLUDE-MIDDLE}.
This rule splits the case in the entailment, identifying formulas that force a shape predicate into either its base or recursive case, due to contradicting pure constraints.
The proof then branches, one branch enforcing the base case, and one enforcing the recursive case. 
For the second condition of NF, \trule{NODE-EX} explicates the fact that given a heap \form{\sheaps}, for any \form{l \in \dom({\sheaps})}, then \form{l\neq \nil}.
For the third condition of NF, \trule{NODEs-EX} explicates the semantics of the separating conjunction \form{\sepconj}: given a heap \form{\sheaps}, for any \form{l_1, l_2 \in \dom({\sheaps})}, then \form{l_1\neq l_2}.
For the fourth condition of NF, while \trule{LIDENT=} discards redundant equalities, \trule{LS-LBASE}, \trule{SLS-LBASE}, \trule{TREE-LBASE} and \trule{STREE-LBASE} concretises the inductive predicates with its base case prior to applying any corresponding substitution.
While the fifth condition of NF could be checked syntactically, the sixth would be performed using an SMT solver such as Z3 \cite{deMoura:TACAS:2008}.

Transforming the bi-abduction problem into it's normalised form is a key stage of our technique.
By normalising the entailment, ambiguity in the entailment is eliminated and any potential contradictions are identified.
This precise representation is critical to the effectiveness of our rules.


%% file: matchsub.tex
\section{Matching and Subtraction}
\label{sec.matchsub}
\begin{figure*}[t!]
\begin{center}
	\[
		\begin{array}{c}
			\rulename{RIDENT=}\\
			\AxiomC{$\D \sepconj [M] \judge \D'\sepconj [F]$}
		    \UnaryInfC{$\D \sepconj [M] \judge \D'\conj E{=}E \sepconj [F]$}
			\DisplayProof
		\end{array}
		\begin{array}{c}
			\rulename{*{-}INTRODUCTION}\\
		    \AxiomC{$\Sigma \sepconj [M_1] \judge  \Sigma' \sepconj [F_1]$}
			\AxiomC{$\Pi \conj \Sigma_1 \sepconj [M_2] \judge  \Pi' \conj \Sigma_2 \sepconj [F_2]$}
		    \BinaryInfC{$\Pi \conj \Sigma \sepconj \Sigma_1 \sepconj [M_1 \sepconj M_2] \judge \Pi' \conj \Sigma' \sepconj \Sigma_2 \sepconj [F_1 \sepconj F_2]$}
			\DisplayProof
		\end{array}
	\]
	\[
		\begin{array}{c}
			\rulename{HYPOTHESIS}\\
			\AxiomC{$\Pi \conj \Sigma \sepconj [M] \judge  \Pi'' \conj \Sigma' \sepconj [F]$}
			\UnaryInfC{$\Pi \conj \Sigma \sepconj [M] \judge \Pi' \conj  \Pi'' \conj \Sigma' \sepconj [F]$}
			\DisplayProof \\
			(\text{where } \form{\Pi \Rightarrow \Pi'})
		\end{array}
		\begin{array}{c}
			\rulename{LS{-}BASE}\\
			\AxiomC{$\D \sepconj [M] \judge \D'\sepconj [F]$}
			\UnaryInfC{$\D \sepconj [M] \judge \D'\sepconj ls(E,E)\sepconj [F]$}
			\DisplayProof
		\end{array}
		\begin{array}{c}
			\rulename{TREE-BASE}\\
			\AxiomC{$(\D) \sepconj [M] \judge \D'\sepconj [F]$}
			\UnaryInfC{$\D \sepconj [M] \judge \D'\sepconj tree(\nil)\sepconj [F]$}
			\DisplayProof
		\end{array}
	\]
	\[
		\begin{array}{c}
			\rulename{SLS{-}BASE}\\
			\AxiomC{$\D \sepconj [M] \judge \D'[V'/V]\sepconj [F]$}
			\UnaryInfC{$\D \sepconj [M] \judge \D'\sepconj sls(E,V,V',E)\sepconj [F\conj V=V']$}
			\DisplayProof
		\end{array}
		\begin{array}{c}
			\rulename{STREE-BASE}\\
			\AxiomC{$\D \sepconj [M] \judge \D'[V'/V]\sepconj [F]$}
			\UnaryInfC{$\D \sepconj [M] \judge \D'\sepconj stree(\nil,V,V')\sepconj [F \conj V{=}V']$}
			\DisplayProof
		\end{array}
	\]
	
	\[
		\begin{array}{c}
			\rulename{SLS-REC (Node)}\\
			\AxiomC{$  \Delta \sepconj [M] \judge   \colorbox{gray!30}{$\exists V_2.~ V_1\lteq V_2$} \conj \Delta' \sepconj sls(E_2,V_2,V_3,E_3) \sepconj  [F]$}
			\UnaryInfC{$\Delta \sepconj E_1\pointsto[E_2,V_1] \sepconj [M] \judge \Delta' \sepconj sls(E_1,V_1,V_3,E_3)\sepconj  [F]$}
			\DisplayProof \\(\text{where }  E_1\pointsto[E_2,V_1] \not\in \D')
		\end{array}		
		\begin{array}{c}
			\rulename{LS-REC (Node)}\\
			\AxiomC{$ \Delta \sepconj [M] \judge  \Delta' \sepconj ls(E_2,E_3) \sepconj  [F]$}
			\UnaryInfC{$\Delta \sepconj E_1\pointsto[E_2] \sepconj [M] \judge \Delta' \sepconj ls(E_1,E_3) \sepconj  [F]$}
			\DisplayProof \\(\text{where }  E_1\pointsto[E_2] \not\in \D')
		\end{array}
	\]
	\[
           \begin{array}{c}
			\rulename{R-EX}\\
			\AxiomC{$\Delta \sepconj [M] \judge \Delta'[V'/V]  \sepconj  [F]$}
			\UnaryInfC{$\Delta \sepconj [M] \judge \exists V.~ \Delta'  \sepconj  [F]$}
			\DisplayProof \\(\text{where }  V' \in \FV(\D))
		\end{array}
		\begin{array}{c}
			\rulename{TREE-REC (Node)}\\
			\AxiomC{$\Delta \sepconj [M] \judge \Delta' \sepconj tree(l)\sepconj tree(r) \sepconj  [F]$}
			\UnaryInfC{$\Delta \sepconj E\pointsto[l,r] \sepconj [M] \judge \Delta' \sepconj tree(E) \sepconj  [F]$}
			\DisplayProof \\(\text{where }  E\pointsto[l,r] \not\in \D')
		\end{array}
	\]
	\[
		\begin{array}{c}
			\rulename{STREE-REC (Node)}\\
			\AxiomC{$ \Delta \sepconj [M] \judge  \colorbox{gray!30}{$V_1\lteq V_2 \conj V_2\lteq V_3$} \conj \Delta' \sepconj stree(l,V_1,V_2)\sepconj stree(r,V_2,V_3) \sepconj  [F]$}
			\UnaryInfC{$\Delta \sepconj E_1\pointsto[l,r,V_2] \sepconj [M] \judge \Delta' \sepconj stree(E_1,V_1,V_3) \sepconj  [F]$}
			\DisplayProof \\(\text{where }  E_1\pointsto[l,r,V_2] \not\in \D')
		\end{array}
	\]
	\[
		\begin{array}{c}
			\rulename{LS{-}REC}\\
			\AxiomC{$\D  \sepconj [M] \judge \D' \sepconj ls(E_2,E_3) \sepconj [F]$}
			\UnaryInfC{$\D \sepconj ls(E_1,E_2) \sepconj [M] \judge \D'\sepconj ls(E_1,E_3)\sepconj [F]$}
			\DisplayProof \\
			(\text{where } \form{E_3} \text{ is dangling } \&~ ls(E_1,E_2) \not\in \D')
		\end{array}
		\begin{array}{c}
			\rulename{SLS{-}REC}\\
			\AxiomC{$\D  \sepconj [M] \judge \colorbox{gray!30}{$\exists V_2.~V'{\leq}V_2$} \conj \D' \sepconj sls(E_2,V_2,V_3,E_3)\sepconj [F]$}
			\UnaryInfC{$\D \sepconj sls(E_1,V_1,V',E_2) \sepconj [M] \judge \D'\sepconj sls(E_1,V_1,V_3,E_3)\sepconj [F]$}
			\DisplayProof \\
			(\text{where } \form{E_3} \text{ is dangling }  \&~ sls(E_1,V_1,V',E_2) \not\in \D')
		\end{array}
	\]
\caption{Subtraction Rules for Bi-Abduction}
\label{fig.ent.sub}
\end{center}
\end{figure*}

We will now discuss the process of matching and subtraction in our technique, alongside the rules underpinning this design. 
At it's core, this second phase of the search is aimed at identifying and resolving equivalent and overlapping segments of the entailment, aiming to reduce both sides of the entailment to \texttt{emp}.
The subtraction rules, detailed in Figure \ref{fig.ent.sub}, aim to achieve this goal in one of three ways, with each rule grouped according to manner in which the rule operates.

The first group consists of decision rules, \trule{EMP} and \trule{IDENT} and are defined as follows:

\[
\begin{array}{c}
	\rulename{EMP}\\
	\AxiomC{}
	\UnaryInfC{$\Pi \conj \emp \sepconj [\true \conj \emp] \judge \true \conj \emp \sepconj [\Pi \conj \emp]$}
	\DisplayProof
\end{array}
\]

\[
\begin{array}{c}
	\rulename{IDENT}\\
	\AxiomC{}
	\UnaryInfC{$\D \sepconj [\true \conj \emp] \judge \D \sepconj [\true \conj \emp]$}
	\DisplayProof
\end{array}
\]

Both of these axioms are applied only in a situation where the search can be ended.
\trule{EMP} is applied whenever the antecedent contains only pure constraints and the consequent is empty, carrying forwards those pure terms into the frame, preserving those additional constraints for future use.
\trule{IDENT} is applied only when the antecedent and the consequent are identical, matching and removing all terms, and ending the search.
As both sides are identical in this case, no additional terms are inferred, leaving the frame and anti-frame unchanged.

The second group describes rules that seek to reduce the \textit{size} of the entailment (Definition \ref{def.size}, Section \ref{sec.sound}) by subtracting matching sub-formulae from each side, and are presented in Figure \ref{fig.ent.sub}. 
Two rules, \trule{RIDENT=} and \trule{HYPOTHESIS} focus on removing pure formulae from the entailment via elimination of trivial and \draftHL{matching }expressions, respectively. 
While \trule{RIDENT=} simply removes the trivial term, \trule{HYPOTHESIS} identifies a sub-formula of pure terms from the antecedent that implies a sub-formula in the consequent, removing the sub-formula and maintaining the terms in the antecedent. 
\trule{*-INTRODUCTION} operates in a similar manner to \trule{HYPOTHESIS}, identifying and separating out spatial terms that entail each other, directly or otherwise.
Unlike \trule{HYPOTHESIS}, however, \trule{*-INTRODUCTION} creates a new branch in the proof, aiming to demonstrate the entailment holds explicitly, and allowing for indirect equivalences as a result.
We also include a rule to concretise existentially-quantified terms in the consequent, \trule{R-EX}. This rule specifically replaces the existential variable in the consequent with a free variable selected from the antecedent, with the intent of allowing for further subtraction of terms. 

The bulk of the remaining rules in this group are specialised unfolding and subtraction rules, targeted at specific shape predicates.
\trule{BASE} rules identify and eliminate shape predicates that are equivalent to \texttt{emp}, preserving any additional relations enforced by the definition. 
Rules \trule{REC} and \trule{REC (Node)} aim to identify, unfold and subtract any overlapping spatial formulae, preserving any implicit constraints and maintaining any unaffected fragments, where necessary.
\trule{REC} rules target scenarios with two overlapping spatial predicates, and \trule{REC (Node)} rules target a points-to term that overlaps with a specific shape predicate. 
As an example, the rule \trule{SLS-REC} would be applied to scenarios in which an SLS predicate in the antecedent is found to have the same head pointer as an SLS predicate in the consequent, as in $sls(x,i,j,y) \mayentail sls(x,i,k,z)$.
Rules targeting other data structures would be utilised in the same way. 
\draftNote{Want to add something about TREE-REC.}
\draftNote{An important point regarding the lack of \trule{REC} rules targeting $tree$ and $stree$ predicates is that REC rules are a specialised case of REC (Node), only permissible when an end-point is known. As Trees do not have a sinular end-point or a collection of leaf nodes, such rules cannot be applied effectively. As a result, only Rec (Node) rules exist for tree predicates in our system.}

The third group consists of the two following rules:

\[
	\begin{array}{c}
	\rulename{LS-GENERALIZE}\\
	\AxiomC{$V{\leq}V' \conj \Delta \sepconj [M] \judge \Delta'\sepconj ls(E_2,E_3) \sepconj [F]  $}
	\UnaryInfC{$\Delta \sepconj sls(E_1, V, V', E_2) \sepconj [M] \judge \Delta' \sepconj ls(E_1,E_3) \sepconj [F]$}
	\DisplayProof \\(\text{where }  sls(E, V, V', E') \not\in \D')
	\end{array}
\]

\[
	\begin{array}{c}
	\rulename{tree-GENERALIZE}\\
	\AxiomC{$V{\leq}V' \conj \Delta  \sepconj [M] \judge \Delta' \sepconj [F] $}
	\UnaryInfC{$\Delta \sepconj stree(E, V, V') \sepconj [M] \judge \Delta' \sepconj tree(E) \sepconj [F]$}
	\DisplayProof \\(\text{where }  stree(E, V, V') \not\in \D')
	\end{array}
\]
\color{black}
These two rules inductively generalize ordering constraints in the LHS such that the matching between the two could be successful.
In our system, a sorted linked list indirectly entails a basic singly-linked list between the same points due to the fact that by our definition, all sorted lists are simple singly-linked lists at the structural level.
This fact could be easily proved by inductively unfolding both predicates and eliminating the matching nodes.
\trule{LS-GENERALIZE} fires when a sorted list in the antecedent of the entailment shares a head node with a simple singly-linked list on the consequent, and essentially abstracts the ordering information away from the sorted list, leaving only the spatial information; a singly-linked list. \trule{TREE-GENERALIZE} operates in the same manner.

The normalisation and subtraction rules of our system are sufficient to prove that a given \textit{valid} entailment holds.
However, should an entailment be invalid, with either some aspect of the symbolic heap missing or extra, these rules will not be enough, and our search must continue on to our inference rules. 


%% file: inf.tex

\section{Inference}
\label{sec.inf}

\input{biabd_table_rules}

When normalisation and subtraction rules are insufficient to continue the proof search, the search aims to find a applicable rule within our set of inference rules.
The rules within this set aim to identify a relevant fragment of the anti-frame or frame, determined by the current state of the entailment, that will satisfy some extra or missing term that cannot be resolved by the previous rules.
These identified fragments are introduced into the relevant part of the entailment, and are matched against the existing terms, eliminating such matches where possible.
All of the identified fragments are maintained during the search, with any fragment of the frame being propagated forwards and the fragments of the anti-frame collected and propagated backwards, producing the final specification.
As with the other rule groups, once an appropriate rule match is found, the search will loop back to the beginning of the rule search; this approach ensures that the inference rules are applied only when absolutely necessary.

The proof rules for inference are shown in Figure \ref{biabd.fig.rules}.
\trule{INF{-}PURE} is the final axiom in our core system, and is applied whenever both sides of the entailment consist of pure terms only. 
\trule{INF-PURE} identifies that all remaining pure terms do not have a direct or indirect match, and moves all remaining pure terms in the antecedent into the frame, and all remaining pure terms in the consequent into the anti-frame, ending the search.
Important to note is that the side-condition of the rule implies that we will never infer a trivial pure constraint.
\trule{INF{-}\form{\pointsto}} infers the unification of constraints over fields such that two points-to predicates could be matched.
This new equality is introduced into the anti-frame and antecedent, and the now matched predicates are removed. 
\trule{INF{-}SLS} and \trule{INF{-}LS} identify overlap between an SLS or LS predicate in the antecedent and subtract it from the entailment.
The rules also infer the constraints required to allow for this subtraction, adding any inferred terms to the anti-frame and antecedent. 
In the case of \trule{INF-SLS}, this may also include the ordering constraints.

\draftNote{Unsure whether this section should go here. A more general discussion about the rules targeting the sorted list predicates.}
In order to maintain the validity of the sorted list predicates, both \trule{SLS-REC*} and \trule{INF{-}SLS} take additional steps to ensure that the order of the sorted list is preserved throughout the application of the rule.
In this case, this means the identification of a suitable ordering constraint between the sorted list and other spatial predicates.
Unlike the inequality between the two end-points, however, this ordering constraint is not abduced from the application of the rule, as it is implicit inside the sorted list predicate throughout.
Instead, the ordering constraint is simply made explicit by the application of the rule.
For the \trule{INF{-}SLS} rule, the ordering constraint is immediately added to the known symbolic state, working under the assumption that the logical variable $X$ refers to a node with a value field, required for the correctness of the $sls$ predicate that remains.
For \trule{SLS-REC*}, the ordering constraint becomes an additional constraint in the consequent of the entailment, requiring proof that the ordering relation is satisfied in order to prove the overall entailment.
The inference of these ordering constraints may be performed through \trule{INF{-}PURE} afterwards, if needed.
While this may initially seem to be a useful aspect in ensuring the soundness of the \trule{SLS-REC*} and \trule{SLS-*} rules, this identification of formerly implicit ordering constraints has an additional benefit.
In previous works in the area of combined domain bi-abduction, the handling of complex pure terms, such as ordering, was typically undertaken 
in a secondary analysis phase.
By completely integrating the pure constraints into the abduction and entailment rules, we have developed a system in which the complex pure properties are identified \textit{during} the shape analysis, eliminating the need for a subsequent pure analysis following this initial phase.

Finally, \trule{INF{-}MISSING} and \trule{INF{-}EXTRA} identify the missing symbolic heap of the antecedent and add them to the anti-frame and frame, respectively.
While this is typically necessary, bulk additions to the anti-frame may result in the identification of overly-complex specifications, should the previous rules not be applied.
The greater precision of the earlier rules is therefore preferred. In these two rules, to avoid inference contradiction, the satisfiability of the solution is required. The solving of this satisfiability may be implemented through existing works like \cite{Le:CAV:2016,Bach:ICSME:2016,Gu:IJCAI:2016,Le:CAV:2017,Xu:CADE:2017}.


%% file: biabd_table_rules.tex
\begin{figure*}[tb]
	\begin{center}
		\[
		\begin{array}{c}
			\rulename{INF{-}PURE}\\
			\AxiomC{}
			\UnaryInfC{$(\Pi \conj \emp)\sepconj [\Pi' \conj \emp] \judge \Pi' \conj \emp \sepconj [\Pi \conj \emp]$}
			\DisplayProof \\
		 	(\text{where } \form{\Pi \wedge \Pi'} \text{ is satisfiable.})
		\end{array}
		\begin{array}{c}
			\rulename{INF{-}\form{\pointsto}}\\
			\AxiomC{$(E_0 \eq E_1 \conj \Delta) \sepconj [M] \judge \Delta' \sepconj [F]$}
			\UnaryInfC{$\Delta \sepconj E\pointsto E_0 \sepconj [E_0 \eq E_1 \conj M] \judge \Delta' \sepconj E\pointsto E_1\sepconj [F] $}
			\DisplayProof
		\end{array}
		\]
		\[
		\begin{array}{c}
			\rulename{INF-LS}\\
			\AxiomC{$E\neq E_1 \conj \Delta \sepconj ls(X, E_1) \sepconj [M] \judge \exists \vecVar{Y}.\Delta'$}
			\UnaryInfC{$\Delta \sepconj ls(E,E_1) \sepconj [E\neq E_1 \conj M] \judge \exists X\vecVar{Y}.\Delta' \sepconj E\pointsto [X]$}
			\DisplayProof	
		\end{array}
		\]
		\[
		\begin{array}{c}
			\rulename{INF-TREE}\\
			\AxiomC{$X \neq \nil \conj \Delta \sepconj tree(E) \sepconj tree(F) \mayentail \exists \vecVar{Y} \Delta'$}
			\UnaryInfC{$\Delta \sepconj tree(X) \sepconj [X \neq \nil \conj M] \mayentail \exists X\vecVar{Y}\Delta' X\pointsto[E,F]$}
			\DisplayProof
		\end{array}
		\]
		\[
		\begin{array}{c}
			\rulename{INF-SLS}\\
			\AxiomC{$E_1\neq E_2 \conj V_1 \lteq V' \conj \Delta \sepconj sls(X,V',V_2,E_2) \sepconj [M] \judge \Delta' \sepconj [F]$}
			\UnaryInfC{$\Delta \sepconj sls(E_1,V_1,V_2,E_2) \sepconj [E_1\neq E_2 \conj V_1 \lteq V' \conj M] \judge \Delta' \sepconj E_1\pointsto [X,V_1] \sepconj [F]$}
			\DisplayProof
		\end{array}
		\]
		\[
		\begin{array}{c}
			\rulename{INF-STREE}\\
			\AxiomC{$X \neq \nil \conj V_3\lteq V \conj V\lteq V_4 \conj \Delta \sepconj stree(E,V_1,V_3) \sepconj stree(F,V_4,V_2) \mayentail \exists \vecVar{Y} \Delta'$}
			\UnaryInfC{$\Delta \sepconj stree(X,V_1,V_2) \sepconj [X \nteq \nil \conj V_3\lteq V \conj V\lteq V_4 \conj M] \mayentail \exists X\vecVar{Y}\Delta' X\pointsto[E,F,V]$}
			\DisplayProof
		\end{array}
		\]
		\[
		\begin{array}{c}
			\rulename{INF{-}MISSING}\\
			\AxiomC{$\Delta \sepconj [M] \judge \Delta' \sepconj[F]$}
			\AxiomC{$\Delta \sepconj Q(E,E') \notentail \false$}
			\BinaryInfC{$\Delta \sepconj [M \sepconj Q(E,E')] \judge \Delta' \sepconj Q(E,E') \sepconj[F]$}
			\DisplayProof\\
			(\text{where } \form{Q(E,E') \text{ is } op(E)})
		\end{array}
		\begin{array}{c}
			\rulename{INF{-}EXTRA}\\
			\AxiomC{$\Delta \sepconj [M] \judge \Delta' \sepconj[F]$}
			\AxiomC{$\Delta' \sepconj Q(E,E') \notentail \false$}
			\BinaryInfC{$\Delta \sepconj Q(E,E') \sepconj [M] \judge \Delta'  \sepconj[F \sepconj Q(E,E') ]$}
			\DisplayProof\\
			(\text{where } \form{Q(E,E') \text{ is } op(E)})
		\end{array}
		\]
		\caption{Inference Rules for Combined Domain}
		\label{biabd.fig.rules}
	\end{center}
\end{figure*}

%% file: sound.tex
	
\section{Soundness and Termination}\label{sec.sound}
We will now present the soundness and termination proof sketches for our system.
\begin{thm}[Soundness]
Given an antecedent \form{\D_a} and a consequent \form{\D_c}, if our system returns \qt{yes} with an anti-frame \form{M}
and a frame \form{F}, then \form{\D_a \sepconj {M} ~\models~ \D_c \sepconj {F} } holds.
\end{thm}
\proofsketch{
We discuss the soundness of the main proof rules in Section \ref{sec.norm}, Section \ref{sec.matchsub} and Section \ref{sec.inf}. Rules {\trule{BASE}} in Figure \ref{fig.ent.norm} are essentially unfolding rules (i.e., replace an inductive predicate by its definition) prior to pruning the recursive case due to the contradiction of equality constraints in the LHS.
{\trule{NODE{-}EX}} is based on the fact that \form{G(op(E)) \wedge op(E) \Rightarrow E\neq \nil}.
{\trule{NODEs{-}EX}} is based on the fact that \form{G(op_1(E_1)) \wedge G(op_2(E_2)) \wedge op_1(E_1) \sepconj op_2(E_2)  \Rightarrow E_1\neq E_2}.
In Figure \ref{fig.ent.sub}, rules {\trule{BASE}} (reps. {\trule{REC(NODE)}}) are essentially unfolding rules prior to pruning the recursive case (resp. base case).
{\trule{LS-REC}} is based on the fact that \form{ls(E_1,E_2) \sepconj ls(E_2,E_3) \Rightarrow ls(E_1,E_3)}. Similarly, {\trule{SLS-REC}} is based on the fact that \form{V_2{\leq}V_3 \wedge sls(E_1,V_1,V_2,E_2) \sepconj sls(E_2,V_2,V_3,E_3) \Rightarrow sls(E_1,V_1,V_3,E_3)}.
In Figure \ref{biabd.fig.rules}, {\trule{INF-LS}} and {\trule{INF-SLS}} infer the pure constraints for recursive case, unfolding the inductive predicate in the LHS, prune the base case prior to matching and subtracting the points-to predicates.
}

Termination of our system is based on the size of an entailment which is defined as:
\begin{definition}[Size]
\label{def.size}
The size of an entailment \form{{\Pi_a \wedge \Sigma_a} \judge {\Pi_c \wedge \Sigma_c}} is a triple of:
\begin{enumerate}
\item \form{N_p{-}n_p} where \form{N_p} is the maximal number of both points-to predicates and inductive predicates that the RHS of any entailments derived from the entailment may contain, and \form{n_p} is the total number of both points-to predicates and inductive predicates in \form{\Sigma_c}.
\item \form{N_e{-}n_e} where \form{N_e} is the maximal number of both disequalities and non-trivial equalities that the LHS
of any entailments derived from the entailment may contain, and \form{n_e} is the number of both disequalities and non-trivial equalities in \form{\Pi_a}.
\item the sum of the length of \form{\Pi_a \wedge \Sigma_a ~\judge~ \Pi_c \wedge \Sigma_c}, where length is defined in the obvious way taking all simple formulas to have length 1.
\end{enumerate}
\end{definition}

\begin{thm}[Termination]
Our system terminates.
\end{thm}

\proofsketch{
\form{N_p} and \form{N_e} are bound.
Moreover, every rule reduces the size: each rule in Fig. \ref{fig.ent.norm} increases either \form{n_p} or \form{n_e};
each rule in Fig. \ref{fig.ent.sub} reduces the length (the third component); each rule in Fig. \ref{biabd.fig.rules} reduces the length.
}

%% file: implement.tex

\section{Implementation}
	\label{implement}

	Our proof technique has been prototyped as an instantiation of a modified version of the Cyclist library\footnote{http://www.cyclist-prover.org}.
	While the library is largely intended to be used to identify \textit{Cyclic} proofs of programs \cite{Brotherston:POPL:2008,Brotherston:ASPLS:2012,Brotherston:SAS:2014}, the underlying representation of separation logic is quite robust and the proof-based approach used in the library is very similar to our own.
	
	Our implementation is broadly-speaking a modification of one of the existing instantiations of the Cyclist library, a separation logic entailment prover, \texttt{sl\_prove}.
	While the general usage and structure of our prototype is very similar to that of the original instantiation, there were two main changes made to the tool.
	Firstly, a number of modifications were introduced to enable the retrieval and validation of the frames and anti-frames produced by the technique. 
	The proof rules themselves were implemented as a replacement of the rule-set utilised by the original tool.
	Secondly, a new proof searching algorithm was implemented as three aforementioned phases: unfolding and normalization, matching and subtracting, and inference.
	In addition to these modifications, a number of additions and changes were made to the underlying separation logic representation.
	These changes were primarily focussed on the extension of the initial fragment to include support for ordering constraints, a basic degree of support for numerical values, and a number of critical supporting functions and behaviours that allows for such support to be effectively utilised.
	
	Wherever possible, these changes were performed in a manner which had little to no effect on the other tools in the library, allowing for the other instantiations of the tool to be used without issues.
	While the bulk of these changes were made in this way, either preserving older behaviour or replacing code in a way that preserved the signatures, some changes were to distinct from the original code for this to be effectively achieved.
	In these cases, new modules were created, generally similar in design and function, but distinct due to the enhanced capabilities.
	As a result, the other existing instantiations of the Cyclist library should still operate correctly. 

	In order to more effectively and efficiently support an automated implementation, a small number of conditions were applied to the implementation of the rules.
	First, a number of restrictions on rule applications were introduced in order to prevent irrelevant or unhelpful applications. 
	As an example, the existential renaming rule \trule{R-EX} can only be applied in the implementation if the identified renaming directly enables the use of a subsequent subtraction rule, preventing spurious renaming of existential variables.
	Secondly, both \trule{HYPOTHESIS} and \trule{*-INTRO} rules were reduced in scope to elimination of singular matching formulas.
	While an unrestricted version was implemented and tested, the impact on the efficiency of the implementation was significant, slowing many analyses down to near-unusable speeds, and in some cases exhausting all available memory.
	The restricted variant currently used in our implementation avoids these issues, and while less comprehensive, does not seem to have an impact on the results.
	Finally, \trule{EXCLUDE-MIDDLE} was restricted to only splitting on cases using the spatial terms within given predicate, limiting the branching only to specific empty or non-empty cases.
	This restriction in particular was due to the significant performance cost of permitting any and all case-splits to be made via \trule{EXCLUDE-MIDDLE}.
	These restrictions help to gain a significant improvement in efficiency over an unrestricted implementation.

	In addition to the restrictions on specific rules, a number of additional rules were implemented in order to support the implementation. 
	One such example is the \trule{SLS-INV} rule, which makes explicit the ordering invariant of a guarded $sls$ predicate in order to preserve that information though subtraction operations, and a restricted renaming rule \trule{SLS-PARAM-EQ}, which identifies and equates non-spatial parameters in matching predicates. 
	Corresponding rules for trees are also present.
	
	Overall, the prototype does have a number of limitations at present, such as the very general treatment of numerical values and a brittle approach to the implementation of rules that target specific predicates.
	These issues, intended to be resolved in future refinements, do carry a slight negative impact on the effectiveness, stability and general extensibility of the tool, though they do not impact the underlying technique.
	Despite these issues, the initial results indicate that our approach shows promise in the area of combined domain bi-abductive inference.
	These results are presented and discussed below. 
	

%% file: eval.tex


\section{Evaluation}
	\label{sec.eval}

	In order to experimentally evaluate our technique, we applied our prototype tool to a selection of benchmarks utilised in the 2019 SL-COMP competition\footnote{Benchmarks were downloaded from competition GitHub during November 2019 \cite{slcomp:19}.} \cite{Sighireanu:TACAS:2019}.
	Specifically, we performed our evaluation over the benchmarks included in the SL-COMP divisions \texttt{qf\_shls\_entl}, containing entailments restricted to basic singly-linked lists without quantifiers, and a subset of the benchmarks included in the \texttt{qf\_shidlia\_entl} division, specifically those that targeted sorted list segments, also without quantifiers.
	The benchmarks selected from these divisions are all the benchmarks that operate within the symbolic heap fragment currently supported by our technique, and allow us to examine the capabilities of our implementation against a strong, standardised set of benchmarks.
	
	An important point is that while these examples are not strictly designed for use in the evaluation of bi-abductive techniques, there is still a significant overlap: our tool, when applied over the invalid examples of the set, should produce a frame and anti-frame that, if introduced, would make the example \textit{valid}.
	As a result, the main focus of our analysis is the invalid benchmarks, though the valid examples will still be examined for the purposes of execution times and efficiency. 

	\begin{wraptable}{r}{0.33\linewidth}
		\centering
		\begin{tabular}{|c|c|c|}
			\hline
			\textbf{Category} & \textbf{Valid} & \textbf{Invalid} \\\hline
			\multicolumn{3}{|c|}{\texttt{qf\_shls\_entl}}\\\hline
			\texttt{smallfoot} & 23 & 54 \\\hline
			\texttt{clones} & 40 & 60 \\\hline
			\texttt{bolognesa} & 57 & 53 \\\hline
			\texttt{ls} & 6 & 3 \\\hline
			\multicolumn{3}{|c|}{\texttt{qf\_shidlia\_entl}}\\\hline
			\texttt{sls\_join} & 17 & 0 \\\hline			
		\end{tabular}
		\caption{Breakdown of list benchmarks}
		\label{table_bench_summary}
	\end{wraptable}

	In total, we ran our implementation over 313 example entailments: all 296 basic list examples from \texttt{qf\_shls\_entl} (174 valid and 122 invalid) and 17 sorted list examples from \texttt{qf\_shidlia\_entl}, modified to use our definition of sorted lists.
	A more detailed breakdown of the number of entailments from each source can be seen in Table \ref{table_bench_summary}.	
	These benchmarks are each represented in their own dedicated file, with the entailment problem itself defined using a variation of the SMT-LIB format, extended with support for spatial terms \cite{Iosif:SLCOMP:2018}.
	These files are arranged within relevant division directories, alongside an auxiliary information file which summarises properties of the corresponding example, such as the source and the validity.
	For our experimentation, a batch file was created to pass the filepath of each relevant example to the implementation front end (with expected flags), which parses the SMT-LIB file and invokes the main solving process, forwarding the cumulative output to a benchmark file.
	
	For each valid entailment in the set of benchmarks, our implementation identifies either an empty frame and anti-frame or a trivial one, where the final impact of the inferred fragments is essentially negligible.
	For invalid entailments, our implementation either identifies a suitable frame and anti-frame, ensuring that the refined entailment is valid, or determines that there is is no possible solution for that particular benchmark. 
		
	The benchmarking was performed on an Ubuntu 18.04 virtual machine with 8GB of memory, running on an Intel i7-7700 processor.
	A 30s limit on execution time was enforced throughout these experiments.
	
	The benchmarks will be presented in three parts: first, we will present and discuss the results of applying the implementation over examples featuring basic singly-linked lists with no ordering constraints.
	Subsequently, we will discuss the results of the sorted list benchmarks, and finally conclude with the results of applying our technique to examples utilising trees, both simple binary trees as well as binary search trees.

	\subsection{Simple Singly-Linked List Benchmarks}
		The majority of the examples included in the benchmarks were entailments focussed around the manipulation of simple singly-linked lists, likely due to the number and maturity of tools in this specific area.
		As a result of the volume of benchmarks examined, the analysis of these results will be somewhat general, though a small number of specific results from each category will be presented.
		Additional focus is placed on the Smallfoot benchmarks due to the similarities between that tool and our own.
		
		\subsubsection{Smallfoot}
			\input{table_bench_smallfoot}
			Our implementation proved to be highly effective over the Smallfoot benchmarks, typically identifying and evaluating all found solutions in only a few ms of CPU time.
			Our technique was able to solve the vast majority of the examples within this set, either confirming the validity of the entailment or identifying some suitable frame or anti-frame to correct it.
			The execution times and validation percentages of a subset of the benchmarks of this set can be seen in Table \ref{table_spec_smallfoot}.
			
			Many of the benchmarks in the Smallfoot set are essentially equivalences, in which the antecedent and consequent are two descriptions of a symbolic heap.
			As a result, the majority of the work undertaken by the tool is to show those equivalences through match-and-subtract operations, unfolding and eliminating shape predicates from the entailment where possible. 
			Other examples featured aliased pointer fields, requiring the inference of an equality enforcing that alias for the entailment to be satisfied. 
			Of note in this category is the \texttt{smallfoot-vc37} benchmark.
			In this example, the alias is over the pointer itself; resolving aliases such as this is currently beyond the capacity of our implementation, and thus no solution was found for this particular bi-abductive problem.
			
			Another important aspect to note is that several of the examples within this set were made trivial from simplifications applied during the initial reading of the problem from file.
			Due to the chosen representation of equalities in the Cyclist SL library, trivial equalities are naturally filtered out without any input.
			As a result, a small number of the benchmarks featured either antecedents or consequents that were reduced to \texttt{emp} immediately, vastly simplifying the entailment as a whole.
			In this case, the entailment to be solved appears in the simplified form, such as $\emp \mayentail \emp$; while equivalent to the original input, the simplifications applied to reach this point are not detailed in the proof, though the equivalence itself is apparent.
			
			However, the most relevant examples in this set are those that require bi-abduction in order to be effectively resolved.
			A number of the benchmarks of the Smallfoot set feature entailments with aspects of the symbolic heap that are missing or extra within that entailment. 
			In these cases, our technique must identify those fragments, and infer that they form part of the frame or anti-frame.
			Our technique is remarkably effective in these cases, readily identifying fragments of the entailment that are missing or extra in the example, and moving them to the frame or anti-frame as appropriate.
			There are fewer examples in which a spatial fragment is inferred into the anti-frame than into the frame, but these benchmarks are resolved with equal ease. 
			A number of benchmarks and their inferred specifications can be seen in Table \ref{table_spec_smallfoot}.
			
		\subsubsection{LS}
			
			Alongside the larger groups of entailments, there are also a small number of list-targeting examples sourced from \cite{Enea:FMSD:2017}, listed under the group \texttt{ls}.
			This group of entailments features a number of list segments and connecting points-to terms in the antecedent, and requires that the implementation identify the equivalence of those fragments with a single, unified list over the same points defined in the consequent.
			In order to resolve these entailments, the prototype must identify each overlapping fragment of the heap and subtract it from entailment.
			
			Over these examples, our implementation excels, proving each of the 3 valid entailments in less than 5ms CPU time each.
			Indeed, even over the initially invalid examples, our prototype is able to identify suitable frames and anti-frames and complete the proof in typically under 10ms.

		\subsubsection{Clones}	
			\input{table_bench_clones_bolog}
		
			We have also evaluated our implementation's performance over the set of \texttt{Clones} benchmarks.
			This set of examples feature entailments that include a number of unidentified aliases.
			In each of these cases, in order to satisfy the entailment, these aliases must be identified, and an equality representing this alias must be introduced into the anti-frame or frame as needed.
			
			The benchmarks are arranged into 10 groups of similar problems, with each group featuring a different number of terms.
			For the benchmarks in the earliest groups, there are only a small number of terms to be processed, increasing as the benchmarks proceed.
			In each of these groups, examples 01 through 06 are relatively trivial problems featuring empty sides of the entailment or simple equivalences, and are solved either by simplification during the initial construction of the entailment from the benchmark file, or by early applications of one of our axioms.
			The remaining examples in each group are more complex, and are thus the only ones discussed in detail here; a selection of these benchmarks can be found in Table \ref{table_bench_clones}.
			
			The examples within the earlier groups are solved handily by our implementation, with all examples in groups \texttt{clones-01} through \texttt{clones-04} solved in less than one second of CPU time, including validation of results.
			However, beyond this point the size of the entailments begin to have an increasingly large impact on the time required to complete, with \texttt{clones-05} taking over a second for many of the examples, and dramatically increasing from there.
			Indeed, examples in \texttt{clones-06} takes over 10 seconds to complete, and the implementation begins to time out from \texttt{clones-07} onwards.

			In each group, our implementation was able to solve and identify suitable aliases for each instance of \texttt{e7}, \texttt{e9} and \texttt{e10} (timeouts notwithstanding).
			In most instances, this process was undertaken via the application of the \trule{INF-PTO} rule, identifying an alias implied by the shared root of points-to terms in the antecedent and consequent, and applying it to the entailment as a whole.
			This in turn allows for the remainder of the proof to be resolved as standard, matching and subtracting equivalent fragments.
			This process was accomplished reasonably efficiently, though the impact of larger entailments on the overall performance is quite noticeable.
			
			Example \texttt{e8} within each group is a special case.
			Each instance of \texttt{e8} within the \texttt{clones} benchmarks outlines an invalid entailment, in which the alias eventually produces a conflict between parts of the entailment.
			While our implementation does detect this inconsistency, it does not achieve this until candidate specifications are validated, which unfortunately results in only a minimal reduction in execution time.
			It may be possible to accelerate this detection in future works, but is uninvestigated at this time.

			When looking at the breakdown in execution times, it seems that the bulk of the analysis time is spent on the validation of the produced results, becoming increasingly costly as the number of terms and potential solutions increases.
			Indeed, this aspect seems to be common throughout the examples tested here, and while entailments with a small number of terms are solved almost instantly, larger entailments take dramatically more time to complete. 
							
		\subsubsection{Bolognesa}
			The Bolognesa examples consist primarily of a large number of interlinking spatial fragments.
			In order to prove these examples, our technique must identify equivalent fragments of the entailment, often formed of a mix of points-to entailments and list predicates, and eliminate them, with any remaining fragments inferred as a part of the frame or anti-frame.
			A selection of example benchmarks can be found in Table \ref{table_bench_bolognesa}.
			
			The Bolognesa examples were arranged into 20 groups of 10 examples, with each group increasing in the number of terms and expressions included within the entailment. 			
			Much like the \texttt{Clones} examples, our implementation was able to rapidly solve the majority of the smaller examples in very short periods of time; the majority of these earlier problems were solved within only a few seconds of CPU time, validation included.
			However, as in the other examples, as the number of terms increased, the overall execution times increased dramatically.
			While a small number of exceptions were present within the benchmarks, the larger entailments took significantly more time to complete as the number of terms increased, eventually beginning to timeout regularly in the largest examples.
			A few exceptions to this were present, but as in \texttt{clones}, a significant number of the timeouts occurred within the validation stage, with the underlying proof already constructed.
			Indeed, of the 54 timeouts throughout the entire benchmark set, only 20 were due to a timeout in the construction of the proof, with the remaining 34 taking place during the validation.
			
			With that said, when the implementation did complete, the results showed promise: for the majority of the invalid examples, our implementation produced a range of potential solutions for the frame and anti-frame, typically within a reasonable time. \\
		
		Overall, our approach is reasonably efficient and effective over the basic list examples detailed here.
		Despite some issues regarding the scalability of the proof search over large entailments, the search overall was effective, accurate, and fast in the majority of cases, producing useful and accurate specifications. 
		In fact, it seems that the largest factor in the execution time is the verification of the potential results.
		Indeed, throughout the benchmarks, it is apparent that the majority of the execution time for a given entailment is spent upon the validation of the results, not the construction of the underlying proof.  
		While this is clearly not ideal, the speed at which our technique is able to explore and identify all possible solutions for a given entailment does indicate that our approach is effective, and that a more efficient validation mechanism, possibly taking advantage of SMT-based solutions, may help overcome many of these issues; this in particular is a focus for future works.
		
	\subsection{Sorted List Benchmarks}
		\input{table_bench_sls}
		
		The $sls$ benchmarks utilised in this section were taken from the \texttt{qf\_shidlia\_entl} division of SL-COMP. 
		Unfortunately, there are only a small number of such examples within the SL-COMP benchmarks, and so while the results here are a reasonable indicator of the effectiveness of the implementation, they likely do not fully represent the capabilities of our technique.
		
		Regarding the examples themselves, each of the benchmarks examined in this section are valid entailments, each aiming to demonstrate that several indirectly linked sorted lists entail a single combined sorted list over the same range.
		The primary difference between each example is the amount of explicit information regarding the values of each segment.
		For some of these examples, the exact bounds are known, making the ordering between each segment trivial. 
		However, for the majority of the examples, only a selection of the bounds are known, requiring the identification of ordering constraints that will enforce the order between the segments.
		In order to successfully solve these examples, our technique must be able to identify those ordering constraints, and then identify that the multiple linked segments are equivalent to a single sorted list segment between the same two points.
		The results for this section are detailed in Table \ref{table_bench_sls}.
	
		In summary, our technique was able to solve the sorted list examples in very short periods of time, with the longest execution time only just exceeding half a second.
		For each of the cases, implicit or missing ordering constraints were identified, and as in the simple linked-list examples, each valid state of the shape predicates are explored separately, often producing a range of valid specifications upon completion.
		Curiously, the performance of the tool seems to be better over examples with fewer known boundaries, suggesting that the more general symbolic ordering constraints are simpler to identify with our technique than ordering constraints with known values.
		More investigation into this phenomenon may be undertaken at a later date. 
			
		There are some minor issues, however.
		Of particular note is the presence of the trivial numerical terms within the final outputs.
		Terms such as 100$\leq$ 200 are obviously true, but appear within the final output as a side-effect of our simplistic treatment of numerical terms.
		While modifying the implementation to eliminate such terms would not be overly complex, their presence does not cause a significant negative effect, and so has been ignored for now. 
		
		While these examples are relatively simplistic, the ease with with our technique was able to identify the requisite ordering constraints, either explicit or implicit, does show great promise.
		Though not undertaken in this work, extending the combined-domain benchmark set to include more complex examples, including larger numbers of terms and mixing in more standard spatial terms, such as points-to terms and simple list predicates will give a much greater insight into the effectiveness of our technique.
		Additionally, a far wider range of combined-domain benchmarks were included within the SL-COMP set, though most of them targeted data structures or properties that are not currently supported within our technique.
		By extending the fragment supported by our tool, these benchmarks may be added into our evaluation, further testing the effectiveness of our technique.
		Such enhancements will be left for future works.
		
	\subsection{Binary Trees}
		\input{table_bench_tree}
	
		Unlike the other experiments performed in this evaluation, the examples used for exploring the effectiveness of our technique over trees were created by-hand due to difficulties in identifying relevant examples supported by the tool's current implementation.
		These benchmarks were created in the spirit of the Smallfoot benchmarks, aiming to evaluate the implementations performance and capabilities over a range of fairly \qt{primitive} entailments.
		These entailments and their corresponding results can be seen in Table \ref{table_bench_tree}.
				
		For the majority of the examples tested here, it can be seen that the performance of the tool over the tree-based examples is comparable in many ways to the performance over the SL-COMP examples.
		Over these entailments, the tool was able to create a proof tree identifying a number of possible solutions in very short time, typically under 100ms.
		However, as in the previous examples, the amount of time spent validating these examples is significantly larger than the initial identification.
		In particular, entailment 5 spent almost the entirety of its execution time validating the potential solutions, likely due to the inferred anti-frame including a number of additional predicates.
		
		With that said, the performance over the $stree$ entailments does continue to indicate that the core system is very promising.
		The construction of the initial proof tree, including the inference of frames and anti-frames, continues to take only a short time over these simpler entailments, and while the verification continues to be a source of extreme inefficiency, the overall CPU time required remains quite brief.
		Additionally, due to the similarities between the entailments using $tree$ and $stree$ predicates, it can be seen that the inclusion of ordering constraints does not seem to introduce a large cost to performance.
		As in the $sls$ examples previously, the technique reliably identified the omitted ordering constraints necessary for the entailments to hold, as well as identifying any spatial fragments of the frames and anti-frames, with the overall execution time increasing by only a few ms.
			
		In addition to the more standard entailments, we have also included 2 special cases, aiming to examine the performance of the tool when \trule{GENERALIZE} rules are required to solve the bi-abductive problem. 
		In each example, the antecedent features a binary search tree, whereas the consequent describes only simple binary trees. 
		In order to resolve these examples, the technique must identify the relationship between the two predicates and generalize the $stree$ predicate to a simple $tree$, before continuing on to complete the proof.
		As can be seen in the experimental results, the need to apply one of these rules does not have a significant effect on the overall execution time when compared to similar entailments.
		

%% file: table_bench_smallfoot.tex
\begin{table}
	\centering
	\begin{tabular}{|c|c|c|c|}
		\hline
		\textbf{Benchmark:} & \multirow{2}{*}{\textbf{Valid}} & \multirow{2}{*}{\textbf{Time (ms)}} & \textbf{\% of time}\\
		\texttt{Smallfoot} & & & \textbf{validating} \\ \hline
		\texttt{vc09} & Y & $5$ & 53 \\ \hline
		\texttt{vc21} & Y & $< 1$ & 29 \\ \hline
		\texttt{vc28} & Y & $2$ & 61 \\ \hline
		\texttt{vc29} & N & $2$ & 45 \\ \hline
		\texttt{vc33} & N & $5$ & 54 \\ \hline
		\texttt{vc37} & N & $2$ & - \\ \hline
		\texttt{vc40} & N & $3$ & 48 \\ \hline
		\texttt{vc65} & Y & $2$ & 61 \\ \hline
		\texttt{vc70} & N & $9$ & 61 \\ \hline
\end{tabular}
	\caption{Benchmarks for Smallfoot Example}
	\label{table_bench_smallfoot}
\end{table}

%% file: table_bench_clones_bolog.tex
\begin{table*}[!t]
	\centering
	\begin{subtable}{0.49\linewidth}
		\begin{tabular}{|c|c|c|c|}
			\hline
			\textbf{Benchmark:} & \multirow{2}{*}{\textbf{Valid}} & \multirow{2}{*}{\textbf{Time (s)}} & \textbf{\% of time}\\
			\texttt{clones} & & & \textbf{validating} \\ \hline
			\texttt{03\_e07} & N & 0.119 & 59 \\ \hline
			\texttt{03\_e08} & N & 0.118 & No solution  \\ \hline
			\texttt{03\_e09} & N & 0.172 & 64 \\ \hline
			\texttt{03\_e10} & N & 0.123 & 60 \\ \hline
			
			\texttt{04\_e07} & N & 0.619 & 63 \\ \hline
			\texttt{04\_e08} & N & 0.636 & No solution \\ \hline
			\texttt{04\_e09} & N & 0.916 & 69 \\ \hline
			\texttt{04\_e10} & N & 0.62 & 64\\ \hline
			
			\texttt{05\_e07} & N & 2.897 & 67 \\ \hline
			\texttt{05\_e08} & N & 2.97 & No solution \\ \hline
			\texttt{05\_e09} & N & 4.258 & 73 \\ \hline
			\texttt{05\_e10} & N & 2.88 & 67\\ \hline
			
			\texttt{06\_e07} & N & 12.198 & 71 \\ \hline
			\texttt{06\_e08} & N & 12.382 & No solution \\ \hline
			\texttt{06\_e09} & N & 18.013 & 76 \\ \hline
			\texttt{06\_e10} & N & 12.270 & 71 \\ \hline
		\end{tabular}
		\subcaption{Clones Benchmarks}
		\label{table_bench_clones}
	\end{subtable}
 ~
	\begin{subtable}{0.49\linewidth}
		\begin{tabular}{|c|c|c|c|}
			\hline
			\textbf{Benchmark:} & \multirow{2}{*}{\textbf{Valid}} & \multirow{2}{*}{\textbf{Time (s)}} & \textbf{\% of time}\\
			\texttt{bolognesa} & & & \textbf{validating} \\ \hline
			\texttt{10\_e01} & N & 0.385 & 77 \\ \hline
			\texttt{10\_e03} & Y & 0.744 & 77 \\ \hline
			\texttt{10\_e04} & N & 0.740 & 77 \\ \hline
			\texttt{10\_e09} & N & 0.434 & 77 \\ \hline
			\texttt{12\_e01} & Y & 1.052 & 81 \\ \hline
			\texttt{12\_e03} & N & 20.425 & 80 \\ \hline
			\texttt{12\_e04} & N & 1.742 & 80 \\ \hline
			\texttt{12\_e09} & Y & 10.458 & 79 \\ \hline
			\texttt{14\_e01} & Y & 7.02 & 84\\ \hline
			\texttt{14\_e03} & Y & 12.713 & 83\\ \hline
			\texttt{14\_e04} & Y & 2.4 & 82 \\ \hline
			\texttt{14\_e09} & Y & 12.163 & 83 \\ \hline
			\texttt{16\_e01} & Y & 13.797 & 86\\ \hline
			\texttt{16\_e03} & N & 14.325 & 86\\ \hline
			\texttt{16\_e04} & N & 15.155 & 85\\ \hline
			\texttt{16\_e09} & Y & 26.303 & 85\\ \hline
		\end{tabular}
		\subcaption{Bolognesa Benchmarks}
		\label{table_bench_bolognesa}
	\end{subtable}
	\caption{Selection of Benchmarks for Simple Singly-Linked Lists}
	\label{table_bench_ls}
\end{table*}

%% file: table_bench_sls.tex
\begin{table*}[!t]
	\draftNote{Need to add in AF/FR, though that may cause overrun on the table itself.}
	\draftNote{Also issues involving essentially "dead" inferences (w' = 200 * w = 200 and w doesn't appear anywhere else)}
	\draftNote{Also also the issues present in the other benchmarks ($A * B \mayentail C ... !(A) * B \mayentail C => B \mayentail C$)}
	
	\centering
	\begin{tabular}{|c|c|c|}
		\hline
		\textbf{Benchmark} &\textbf{Time (ms)} & \textbf{\% Validation}\\ \hline
		\texttt{sls\_join\_2\_known\_bnd} & 16 & 47 \\ \hline
		\texttt{sls\_join\_2\_no\_cond} & 20 & 62 \\ \hline
		\texttt{sls\_join\_2} & 20 & 63 \\ \hline
		\texttt{sls\_join\_2\_unk\_lower\_bnd} & 18 & 55 \\ \hline
		\texttt{sls\_join\_2\_unk\_upper\_bnd} & 17 & 56 \\ \hline
		\texttt{sls\_join\_3\_2\_cond} & 119 & 67 \\ \hline
		\texttt{sls\_join\_3\_no\_cond} & 50 & 35 \\ \hline
		\texttt{sls\_join\_3} & 120 & 67 \\ \hline
		\texttt{sls\_join\_3\_unk\_lower\_bnd} & 116 & 67 \\ \hline
		\texttt{sls\_join\_3\_unk\_upper\_bnd}  & 71 & 51\\ \hline
		\texttt{sls\_join\_4\_1\_cond\_unk\_both} & 298 & 51 \\ \hline
		\texttt{sls\_join\_4\_1\_cond\_unk\_lower\_bnd} & 217 & 36 \\ \hline
		\texttt{sls\_join\_4\_1\_cond\_unk\_upper\_bnd} & 78 & 35 \\ \hline
		\texttt{sls\_join\_4\_2\_cond\_unk\_lower\_bnd} & 339 & 54 \\ \hline
		\texttt{sls\_join\_4\_2\_cond\_unk\_upper\_bnd} & 445 & 66\\ \hline
		\texttt{sls\_join\_4\_3\_cond} & 570 & 70 \\ \hline
		\texttt{sls\_join\_4\_no\_cond} & 202 & 32 \\ \hline
	\end{tabular}
	\caption{Benchmarks for Sorted List Entailments}
	\label{table_bench_sls}
\end{table*}

%% file: table_bench_tree.tex
\begin{table}[!t]
		
	\centering
	\begin{tabular}{|c|c|c|c|}
		\hline
		\textbf{No.} & \textbf{Entailment} &\textbf{Time (ms)} & \textbf{\% Validation}\\ \hline
		\textbf{1} & \form{x\pointsto[l,r] \mayentail tree(x)} & 2 & 83 \\ \hline
		\textbf{2} & \form{x\pointsto[l,r] \sepconj tree(l) \sepconj tree(r) \mayentail tree(x)} & 9 & 62 \\ \hline
		\textbf{3} & \form{tree(x) \mayentail x\pointsto[l,r] \sepconj tree(l) \sepconj tree(r)} & 7 & 66 \\ \hline
		\textbf{4} & \form{tree(x) \mayentail tree(x) \sepconj tree(y)} & 3 & 70 \\ \hline
		\textbf{5} & \form{x\pointsto[l,r] \sepconj l\pointsto[m,n] \sepconj r\pointsto[s,t] \mayentail tree(x)} & 70 & 98 \\ \hline
		\textbf{6} & \form{tree(x) \sepconj tree(y) \mayentail x\pointsto[i,j] \sepconj y\pointsto[k,l]} & 62 & 76 \\ \hline
		\textbf{7}  & \form{x\pointsto[l,r,i] \mayentail stree(x,i,j)} & 7 & 88 \\ \hline
		\textbf{8} & \form{x\pointsto[l,r,i] \sepconj stree(l,a,h) \sepconj stree(r,j,k) \mayentail stree(x,a,k)} & 18 & 56 \\ \hline
		\textbf{9} & \form{stree(x,a,k) \mayentail x\pointsto[l,r,i] \sepconj stree(l,a,h) \sepconj stree(r,j,k)} & 16 & 63 \\ \hline
		\textbf{10} & \form{stree(x,a,b) \mayentail stree(x,c,d) \sepconj stree(y,i,j)} & 4 & 63 \\ \hline
		\textbf{11} & \form{x\pointsto[l,r,i] \sepconj l\pointsto[m,n,a] \sepconj r\pointsto[s,t,k] \mayentail stree(x,a,k)} & 196 & 97 \\ \hline
		\textbf{12} & \form{stree(x,i,j) \sepconj stree(y,k,l) \mayentail x\pointsto[a,b,i] \sepconj y\pointsto[c,d,k]} & 114 & 74 \\ \hline
		\textbf{13} & \form{stree(x,i,j) \mayentail tree(x)} & 2 & 57 \\ \hline
		\textbf{14} & \form{stree(x,i,k) \mayentail x\pointsto[l,r,i] \sepconj tree(l) \sepconj tree(r)} & 14 & 69 \\ \hline
	\end{tabular}
	\caption{Benchmarks for Tree Entailments}
	\label{table_bench_tree}
\end{table}

%% file: conc.tex


\section{Conclusions and Future Work}\label{concl}

We have presented a novel proof system for the bi-abduction problem in separation logic for sorted lists and binary search trees.
Our system has been designed based on the Unfold-and-Match paradigm: given an entailment, it systematically explores all candidate matching instances of the antecedent and consequent prior to inferring any missing portion.
In order to inspect the effectiveness of the system, we have implemented the technique into a prototype tool based upon the Cyclist library \cite{Brotherston:ASPLS:2012} and tested it over a range of benchmarks from the SL-COMP competition.
While in an early stage, the system does show promise, and further work to improve the system will likely improve it further.

Possible future works include extending the base system to a more expressive fragment, such as the inclusions of more general inductive definitions and additional pure properties such as size, balance and bag or set information.
Our current system is based on the decidable fragment presented in \cite{Berdine:APLAS:2005}, and extending it to a more expressive decidable fragment such as \cite{10.1007/978-3-319-12736-1_17,10.5555/3169142.3169189} is also of interest.
Other possible avenues include using the system to infer error specifications via bi-abduction \cite{Le:NFM:2013} for use in counterexample generation \cite{Pham:FM:2019,Pham:ATVA:2019} and program repair \cite{Bach:ICSME:2016}.
In addition, our system may be embedded inside a program analysis tool e.g. \cite{Calcagno:POPL:2009,Le:CAV:2014,Brotherston:SAS:2014}, allowing our system to be used for full program analysis and specification inference. 


%% file: appendix_table_spec_smallfoot.tex
\begin{sidewaystable}
	\section{Example Inferred Specifications}
	\centering
	\begin{tabular}{|c|c|}
		\hline
		\textbf{Name} & \textbf{Specifications}\\ \hline\hline
		\multirow{4}{*}{\texttt{vc09}}
			& \biab{ls(x1,null) \sepconj ls(x2,null)}{ }{\sepconj ls(x1,null) \sepconj ls(x2,null)}{x1\nteq null \sepconj x2\nteq null \sepconj x1\nteq 2} \\
			& \biab{ls(x1,null) \sepconj ls(x2,null)}{x1=null}{\sepconj ls(x1,null) \sepconj ls(x2,null)}{x1=null \sepconj x2\nteq null} \\
			& \biab{ls(x1,null) \sepconj ls(x2,null)}{x2=null}{\sepconj ls(x1,null) \sepconj ls(x2,null)}{x2=null \sepconj x1\nteq null} \\
			& \biab{ls(x1,null) \sepconj ls(x2,null)}{x1=null\sepconj x2=null}{\sepconj ls(x1,null) \sepconj ls(x2,null)}{x1=null \sepconj x2=null} \\ \hline
		\texttt{vc21} & \biab{x1\nteq null \sepconj ls(x1, null)}{true\conj emp}{ls(x1,null)}{x1\nteq null} \\ \hline
		\multirow{2}{*}{\texttt{vc28}}
			& \biab{x1\nteq null \sepconj x2\nteq null \sepconj x2\pointsto null \sepconj ls(x1,x2)}{true\conj emp}{ls(x1,null)}{x1\nteq null \sepconj x2\nteq null \sepconj x1\nteq x2} \\
			& \biab{x1\nteq null \sepconj x2\nteq null \sepconj x2\pointsto null \sepconj ls(x1,x2)}{x1=x2}{ls(x1,null)}{x1=x2 \sepconj x1\nteq null \sepconj x2\nteq null} \\ \hline
		\multirow{2}{*}{\texttt{vc29}}
			& \biab{\Pi \sepconj x2\pointsto x1 \sepconj ls(x1,x2)}{x1=x3}{x2\pointsto x3 \sepconj ls(x3,x2)}{\Pi \sepconj x1=x3 \sepconj x2\nteq x3} \\
			& \biab{\Pi \sepconj x2\pointsto x1 \sepconj ls(x1,x2)}{x1=x3 \sepconj x2=x3}{x2\pointsto x3 \sepconj ls(x3,x2)}{x1=x3 \sepconj x2=x3 \sepconj \Pi} \\ \hline
		\multirow{2}{*}{\texttt{vc33}}
			& \biab{\Pi \sepconj x1\pointsto x2 \sepconj x3\pointsto x1 \sepconj ls(x2,x3)}{x1=x4}{x3\nteq x4 \sepconj x3\pointsto x4 \sepconj ls(x4,x3)}{\Pi \sepconj x1=x4  \sepconj x2\nteq x3}  \\
			& \biab{\Pi \sepconj x1\pointsto x2 \sepconj x3\pointsto x1 \sepconj ls(x2,x3)}{x1=x4 \sepconj x2=x3}{x3\nteq x4 \sepconj x3\pointsto x4 \sepconj ls(x4,x3)}{\Pi \sepconj x1=x4 \sepconj x2=x3}  \\ \hline
		\texttt{vc37} & \texttt{INVALID} \\ \hline
		\multirow{2}{*}{\texttt{vc40}}
			& \biab{\Pi \sepconj x2\pointsto x3 \sepconj ls(1,null) \sepconj ls(x3,null)}{ }{ls(x1,null) \sepconj ls(x3,null)}{\Pi \sepconj x3\nteq null \sepconj x2\pointsto x3} \\ 
			& \biab{\Pi \sepconj x2\pointsto x3 \sepconj ls(1,null) \sepconj ls(x3,null)}{x3 = null}{ls(x1,null) \sepconj ls(x3,null)}{\Pi \sepconj x3=null \sepconj x2\pointsto x3} \\ \hline
		\multirow{2}{*}{\texttt{vc65}}
			& \biab{\Pi \sepconj x1\pointsto x2 \sepconj ls(x2,null)}{true\conj emp}{ls(x1,null)}{\Pi \sepconj x2\nteq null} \\
			& \biab{\Pi \sepconj x1\pointsto x2 \sepconj ls(x2,null)}{x2=null}{ls(x1,null)}{\Pi \sepconj x2=null} \\ \hline
		\multirow{4}{*}{\texttt{vc70}}
			& \biab{\Pi \sepconj x1\pointsto x2 \sepconj ls(x2,null) \sepconj ls(x3,null)}{ }{x1\pointsto x2 \sepconj ls(x2,null) \sepconj ls(x1,x1)}{\Pi \sepconj x2\nteq null \sepconj x3\nteq null \sepconj x2\nteq x3 \sepconj ls(x3,null)} \\
			& \biab{\Pi \sepconj x1\pointsto x2 \sepconj ls(x2,null) \sepconj ls(x3,null)}{x3=null}{x1\pointsto x2 \sepconj ls(x2,null) \sepconj ls(x1,x1)}{\Pi \sepconj x2\nteq null \sepconj x3=null} \\
			& \biab{\Pi \sepconj x1\pointsto x2 \sepconj ls(x2,null) \sepconj ls(x3,null)}{x2=null}{x1\pointsto x2 \sepconj ls(x2,null) \sepconj ls(x1,x1)}{\Pi \sepconj x2 = null \sepconj x3\nteq null \sepconj ls(x3,null)} \\
			& \biab{\Pi \sepconj x1\pointsto x2 \sepconj ls(x2,null) \sepconj ls(x3,null)}{x2=null \sepconj x3=null}{x1\pointsto x2 \sepconj ls(x2,null) \sepconj ls(x1,x1)}{\Pi \sepconj x2 = null \sepconj x3 = null} \\ \hline
		
\end{tabular}
	\caption{Inferred Specifications for Smallfoot Benchmark}
	
	{\small\textcolor{blue}{Blue} text indicates inferred anti-frames, \textcolor{red}{Red} text indicates inferred frames. $\Pi$ is used as a shorthand for pure constraints from the initial entailment that are not relevant to the inference.}
	\label{table_spec_smallfoot}
\end{sidewaystable}

